\documentclass[prb,aps,twocolumn,superscriptaddress,showpacs]{revtex4-1}

\makeatletter
\renewcommand*{\@fnsymbol}[1]{\ensuremath{\ifcase#1\or \dagger\or \dagger\or \ddagger\or
		\mathsection\or \mathparagraph\or \|\or **\or \dagger\dagger
		\or \ddagger\ddagger \else\@ctrerr\fi}}
\makeatother
\usepackage{color}
\usepackage{graphicx}
\usepackage{epstopdf}
\usepackage{lineno}

\usepackage{float}
\usepackage{lipsum}
\usepackage{amsmath} 
\usepackage{amssymb}	
\usepackage{graphicx} 
\usepackage{color}
\usepackage{hyperref} 
\usepackage{anysize}
\usepackage{hhline,multirow,float}
\usepackage{makecell}
\usepackage{array}
\usepackage{dcolumn}
\newcolumntype{?}{!{\vrule width 1pt}}
\usepackage{color}

\definecolor{model1}{RGB}{0,0,0}
\definecolor{model2}{RGB}{255,0,0}
\definecolor{model3}{RGB}{135,81,81}
\definecolor{model4}{RGB}{0,176,80}
\definecolor{model5}{RGB}{255,165,0}
\definecolor{model6}{RGB}{128,128,0}
\definecolor{model7}{RGB}{0,0,255}

\usepackage{multirow}
\usepackage[table]{xcolor}
\usepackage{capt-of}
\usepackage{wasysym}

\marginsize{1.5 cm}{1.5 cm}{0.5 cm}{0.5 cm}

\usepackage{etoolbox} 
\usepackage{lipsum} 

\makeatletter
\appto{\appendix}{%
	\@ifstar{\def\theequation@prefix{A.}}%
	{}%
}
\makeatother

\begin{document}
\newcommand\bbone{\ensuremath{\mathbbm{1}}}
\newcommand{\ul}{\underline}
\newcommand{\bp}{{\bf p}}
\newcommand{\vl}{v_{_L}}
\newcommand{\vc}{\mathbf}
\newcommand{\be}{\begin{equation}}
\newcommand{\ee}{\end{equation}}
\newcommand{\bk}{{{\bf{k}}}}
\newcommand{\bK}{{{\bf{K}}}}
\newcommand{\cE}{{{\cal E}}}
\newcommand{\bQ}{{{\bf{Q}}}}
\newcommand{\br}{{{\bf{r}}}}
\newcommand{\bg}{{{\bf{g}}}}
\newcommand{\bG}{{{\bf{G}}}}
\newcommand{\hbr}{{\hat{\bf{r}}}}
\newcommand{\bR}{{{\bf{R}}}}
\newcommand{\bq}{{\bf{q}}}
\newcommand{\hx}{{\hat{x}}}
\newcommand{\hy}{{\hat{y}}}
\newcommand{\hd}{{\hat{\delta}}}
\newcommand{\bea}{\begin{eqnarray}}
\newcommand{\eea}{\end{eqnarray}}
\newcommand{\ra}{\rangle}
\newcommand{\la}{\langle}
\renewcommand{\tt}{{\tilde{t}}}
\newcommand{\upa}{\uparrow}
\newcommand{\dna}{\downarrow}
\newcommand{\bS}{{\bf S}}
\newcommand{\vS}{\vec{S}}
\newcommand{\dg}{{\dagger}}
\newcommand{\pdg}{{\phantom\dagger}}
\newcommand{\tphi}{{\tilde\phi}}
\newcommand{\cf}{{\cal F}}
\newcommand{\ca}{{\cal A}}
\renewcommand{\ni}{\noindent}
\newcommand{\ct}{{\cal T}}
\newcommand{\brf}{\bar{F}}
\newcommand{\brg}{\bar{G}}
\newcommand{\jeff}{j_{\rm eff}}
\newcommand{\cvo}{$\alpha$-CoV$_{3}$O$_{8}$}

\title{Ordered magnetism in the intrinsically decorated $j\rm{_{eff}}$ = $\frac{1}{2}$ \cvo}

\author{P.~M.~Sarte}
\affiliation{School of Chemistry, University of Edinburgh, Edinburgh EH9 3FJ, United Kingdom}
\affiliation{Centre for Science at Extreme Conditions, University of Edinburgh, Edinburgh EH9 3FD, United Kingdom}
\author{A.~M.~Ar\'{e}valo-L\'{o}pez} 
\affiliation{School of Chemistry, University of Edinburgh, Edinburgh EH9 3FJ, United Kingdom}
\affiliation{Centre for Science at Extreme Conditions, University of Edinburgh, Edinburgh EH9 3FD, United Kingdom}
\affiliation{Universit\'{e} Lille 1 Sciences et Technologies, UMR 8181 CNRS, Unit\'{e} de Catalyse et Chimie du Solide `UCCS', 59655 Villeneuve d'ASCQ, France}
\author{M.~Songvilay}
\affiliation{Centre for Science at Extreme Conditions, University of Edinburgh, Edinburgh EH9 3FD, United Kingdom}
\affiliation{School of Physics and Astronomy, University of Edinburgh, Edinburgh EH9 3FD, United Kingdom}
\author{D.~Le}
\affiliation{ISIS Facility, Rutherford Appleton Laboratory, Chilton, Didcot OX11 0QX, United Kingdom}
\author{T.~Guidi}
\affiliation{ISIS Facility, Rutherford Appleton Laboratory, Chilton, Didcot OX11 0QX, United Kingdom}
\author{V. Garc\'{i}a-Sakai}
\affiliation{ISIS Facility, Rutherford Appleton Laboratory, Chilton, Didcot OX11 0QX, United Kingdom}
\author{S.~Mukhopadhyay}
\affiliation{ISIS Facility, Rutherford Appleton Laboratory, Chilton, Didcot OX11 0QX, United Kingdom}
\author{S.~C.~Capelli}
\affiliation{ISIS Facility, Rutherford Appleton Laboratory, Chilton, Didcot OX11 0QX, United Kingdom}
\author{W.~D.~Ratcliff}
\affiliation{NIST Center for Neutron Research, National Institute of Standards and Technology, Gaithersburg, Maryland 20899, USA}
\author{K.~H.~Hong}
\affiliation{School of Chemistry, University of Edinburgh, Edinburgh EH9 3FJ, United Kingdom}
\affiliation{Centre for Science at Extreme Conditions, University of Edinburgh, Edinburgh EH9 3FD, United Kingdom}
\author{G.~M.~McNally}
\affiliation{School of Chemistry, University of Edinburgh, Edinburgh EH9 3FJ, United Kingdom}
\affiliation{Centre for Science at Extreme Conditions, University of Edinburgh, Edinburgh EH9 3FD, United Kingdom}
\affiliation{Max-Planck-Institut f\"{u}r Festk\"{o}rperforschung, D-70569 Stuttgart, Germany}
\author{E.~Pachoud} 
\affiliation{School of Chemistry, University of Edinburgh, Edinburgh EH9 3FJ, United Kingdom}
\affiliation{Centre for Science at Extreme Conditions, University of Edinburgh, Edinburgh EH9 3FD, United Kingdom}
\author{J.~P.~Attfield}
\affiliation{School of Chemistry, University of Edinburgh, Edinburgh EH9 3FJ, United Kingdom}
\affiliation{Centre for Science at Extreme Conditions, University of Edinburgh, Edinburgh EH9 3FD, United Kingdom}
\author{C.~Stock} 
\affiliation{Centre for Science at Extreme Conditions, University of Edinburgh, Edinburgh EH9 3FD, United Kingdom}
\affiliation{School of Physics and Astronomy, University of Edinburgh, Edinburgh EH9 3FD, United Kingdom}

\date{\today}

\begin{abstract}
\indent The antiferromagnetic mixed valence ternary oxide $\alpha$-CoV$_{3}$O$_{8}$ displays disorder on the Co$^{2+}$ site that is inherent to the $Ibam$ space group resulting in a local selection rule requiring one Co$^{2+}$ and one V$^{4+}$ reside next to each other, thus giving rise to an intrinsically disordered magnet without the need for any external influences such as chemical dopants or porous media.  The zero field structural and dynamic properties of \cvo~have been investigated using a combination of neutron and x-ray diffraction, DC susceptibility, and neutron spectroscopy. The low temperature magnetic and structural properties are consistent with a random macroscopic distribution of Co$^{2+}$ over the 16$k$ metal sites.  However, by applying the sum rules of neutron scattering we observe the collective magnetic excitations are parameterized with an ordered Co$^{2+}$ arrangement and critical scattering consistent with a three dimensional Ising universality class.  The low energy spectrum is well-described by Co$^{2+}$ cations coupled $via$ a three dimensional network composed of competing ferromagnetic and stronger antiferromagnetic superexchange within the $ab$ plane and along $c$, respectively. While the extrapolated Weiss temperature is near zero, the 3D dimensionality results in long range antiferromagnetic order at $T\rm{_{N}}~\sim$~19~K.  A crystal field analysis finds two bands of excitations separated in energy at $\hbar \omega$ $\sim$ 5 meV and 25 meV, consistent with a $j\rm{_{eff}}=\frac{1}{2}$ ground state with little mixing between spin-orbit split Kramers doublets.  A comparison of our results to the random 3D Ising magnets and other compounds where spin-orbit coupling is present indicate that the presence of an orbital degree of freedom, in combination with strong crystal field effects and well-separated $j\rm{_{eff}}$ manifolds may play a key role in making the dynamics largely insensitive to disorder.
  \end{abstract}

\maketitle

\section{Introduction}

\indent Introducing disorder into condensed matter systems often suppresses common mean-field phases and transitions in favor of states that exhibit unusual critical properties~\cite{fisher,martin2017,sibille2017,edwards1975,edwards1976,raposo,narayan90:42,savary17:118,altman04:93,weichman08:22,hoyos08:100,vojta06:39}.  Examples of such exotic behavior in insulating systems include the study of quenched disorder through doping in both model magnets~\cite{cowley1972properties,Nieuwenhuizen1999} and liquid crystal systems~\cite{radzihovsky99:60,Bellini01:294,toner90:41}.  While the presence of strong disorder disrupts translational symmetry, often resulting in a glassy phase~\cite{Binder86:58} with long range order destroyed for all length scales, the presence of weak disorder can give rise to phases displaying distinct responses for differing length scales.  For example, in model random field systems near a phase transition, critical thermal fluctuations dominate until the length scale of the order parameter becomes large enough where static terms originating from the induced disorder dominate~\cite{barghathi12:109,vojtabook}.  Examples of new disordered-induced phases include the concept of  ``Bragg glass"~\cite{Giamarchi95:52,klein01:413,hernandez04:92,olsson07:98,li03:90} that were first postulated in the context of flux lattices in superconductors~\cite{gingras96:53,fisher97:64,Giamarchi94:72} where Bragg peaks exist, however other properties reflect a glass type response.  A further example of unusual phases in the presence of disorder is the Griffiths phase~\cite{Griffiths69:23,Bray87:59,mccoy69:23} that was first suggested in the context of Ising ferromagnets, where an ordered local region co-exists within a globally disordered phase. So far, the search for new disordered-induced phases have been limited to introducing disorder by doping in the case of solid state materials, or porous media for liquid crystals~\cite{Bellini01:294,Park02:65,Leheny03:67,Clegg03:67} and quantum fluids~\cite{Glyde18:81,Glyde00:84,ponto95:74,sprague95:75,Chan96:49}. \\
\indent One example of theoretical efforts to understand the effects of quenched disorder on the order parameter near a phase transition is random field theory which relates disorder to the lowering of the dimensionality of the underlying universality class~\cite{imry,imry2}.  Model random magnets~\cite{cowley1972properties,Nieuwenhuizen1999,tabei06:97,silevitch07:448} have played a significant role in the development and validation of such theories with an important example being the dilute Ising antiferromagnets such as Fe$_{x}$Zn$_{1-x}$F$_{2}$~\cite{raposo,slani1999,alvarez2012,rodriguez2007} (Fe$^{2+}$, $L=2$ and $S=2$) and Mn$_{x}$Zn$_{1-x}$F$_{2}$~\cite{Cowley1984,birgeneau87:61} (Mn$^{2+}$, $L=0$ and $S=\frac{5}{2}$).  In these magnets, the random occupancy introduced through doping combined with a magnetic field results in a tunable random field.  While these systems show a competition between static and thermal fluctuations driving magnetic order, the dynamics are largely unaltered by the introduction of weak disorder~\cite{inelastic,Leheny2003} and therefore the magnets with weak quenched disorder have collective dynamics very similar to the parent compounds.  Despite significant interest in the community~\cite{vojta13:1550}, the amount of systems that have been shown to host such exotic phases as described above have been limited, in particular, there are few examples of definitive Bragg glass and Griffiths phases.  In this paper, we discuss a system where disorder is not introduced through doping, but rather is inherit to the crystallographic symmetry and therefore is a situation where magnetic disorder is present despite the presence of structural order.\\
\begin{figure*}[htb!]
	\centering
	\includegraphics[width=1.0\linewidth]{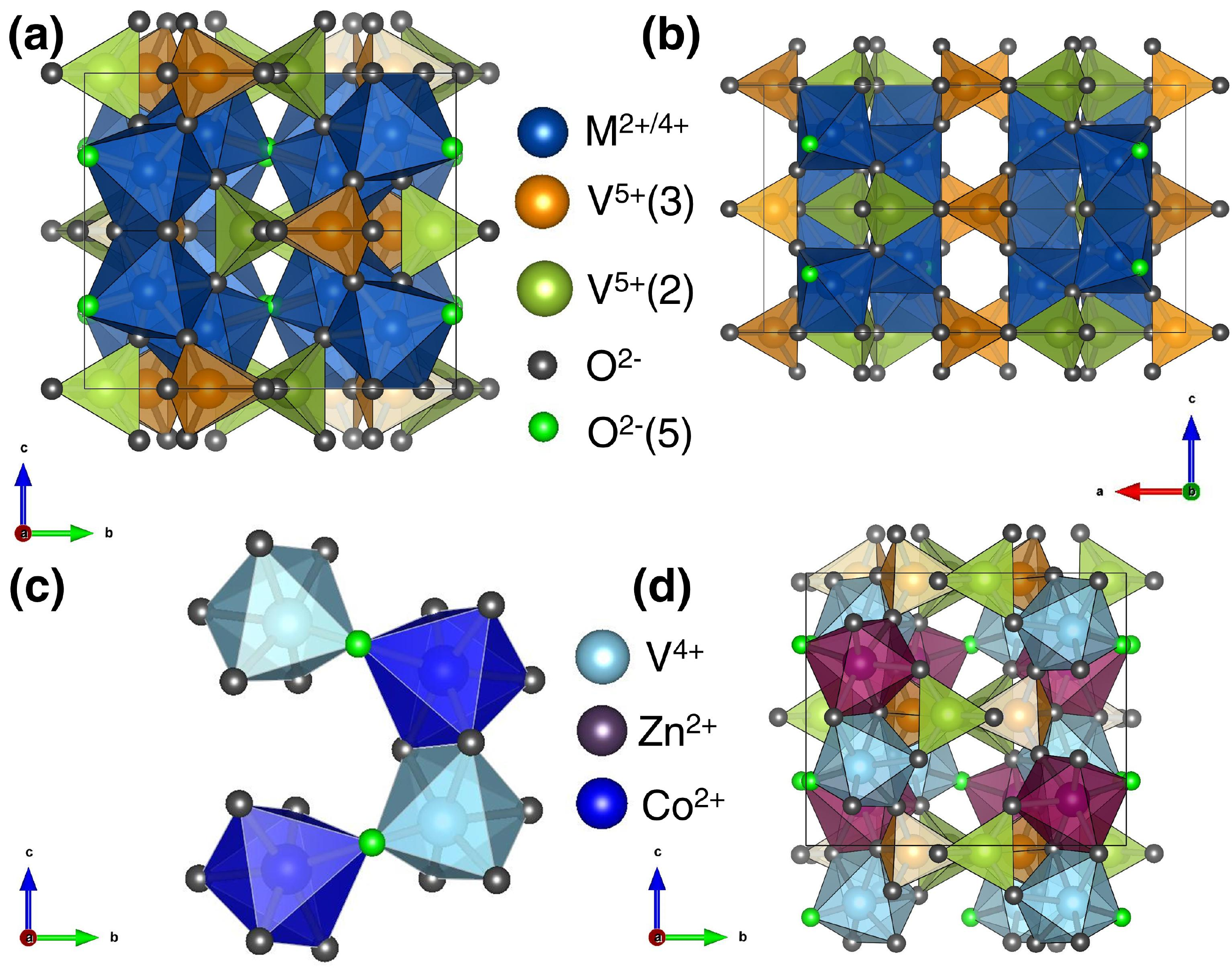}
	\caption{Proposed~\cite{oka} crystal structure of \cvo~($Ibam$, \#72) along the (a) $bc$ and (b) $ac$ planes, consisting of zigzag chains of edge-sharing MO$_{6}$ (M = Co$^{2+}$, V$^{4+}$) octahedra running parallel to $c$ that are interspaced with non-magnetic V$^{5+}$ in tetrahedral (V$^{5+}$(2)) and trigonal bipyramidal (V$^{5+}$(3)) coordination. (c) Local constraint of the $Ibam$ structure. Metal sites opposite of the bridging O(5) must be occupied by one Co$^{2+}$ and one V$^{4+}$, with the O(5) situated closer to the V$^{4+}$ site. (d) Crystal structure of $\alpha$-ZnV$_{3}$O$_{8}$ ($Iba2$, \#45). In contrast to $Ibam$, the $Iba2$ structure consists of an ordered alternating distribution of Zn$^{2+}$ and V$^{4+}$ along the zigzag chains~\cite{znv3o8}.}
	\label{fig:1}
\end{figure*}
\indent In contrast to the disordered systems described above, where the disorder is a consequence of an addition external to the original system ($e.g.$ doping~\cite{Cowley1984,birgeneau87:61,raposo,slani1999,alvarez2012,rodriguez2007}, porous media~\cite{ponto95:74,sprague95:75,Park02:65}, $etc.$), and thus can be finely tuned~\cite{fishman1979random}, the disorder in \cvo~is simply inherent to its $Ibam$ crystal structure. As illustrated in Figs.~\ref{fig:1}(a) and (b), the proposed~\cite{oka} crystal structure of \cvo~consists of zigzag chains of edge-sharing MO$_{6}$ octahedra (M = Co$^{2+}$ and V$^{4+}$) running along $c$. With the exception of a single crystallographic constraint corresponding to a local selection rule requiring that one Co$^{2+}$ and one V$^{4+}$ reside on opposite sides of the O(5) bridging oxygen (Fig.~\ref{fig:1}(c)), the $Ibam$ structure of \cvo~consists of a random distribution of metal cations along the zigzag chains. A combination of the proposed random metal cation distribution with both evidence~\cite{oka} for dominant antiferromagnetic exchange coupling from DC susceptibility and Ising anisotropy due to local axial octahedral distortions and spin-orbit coupling, suggests that \cvo~may represent a potential alternative route for the investigation of disordered-induced physics.  Indeed the study of disorder on electronic structures has found that by introducing correlations, localization~\cite{Anderson58:109} can be suppressed.~\cite{Croy11:82,Moura98:81}  

In this paper, we characterize both the crystal-magnetic structure and fluctuations of \cvo. This paper consists of five sections discussing our results including this introduction and a subsequent section on experimental details. We first present the characterization of the static nuclear-magnetic structure. High resolution single crystal x-ray and neutron diffraction data confirmed both the disordered $Ibam$ crystal structure and the presence of local octahedral distortions. A combination of single crystal magnetic neutron diffraction and single crystal DC susceptibility  identified the presence of ferromagnetic  correlations between Co$^{2+}$ cations within the $ab$ plane, in addition to a dominant antiferromagnetic coupling along $c$. Low energy critical scattering is consistent with 3D Ising behavior attributable to the $j\rm{_{eff}}=\frac{1}{2}$ Co$^{2+}$ ions. However, in contrast to the intrinsically disordered $Ibam$ crystal structure, by applying the first moment sum rule of neutron scattering, we find the excitations are well described by an ordered Co$^{2+}$ arrangement.  We conclude the paper with a section discussing our results in the context of models for disordered magnets and discuss the role of spin-orbit coupling through a comparison of model magnets in a random field.

\section{Experimental Details} 

\begin{figure*}
	\centering
	\includegraphics[width=0.9\linewidth]{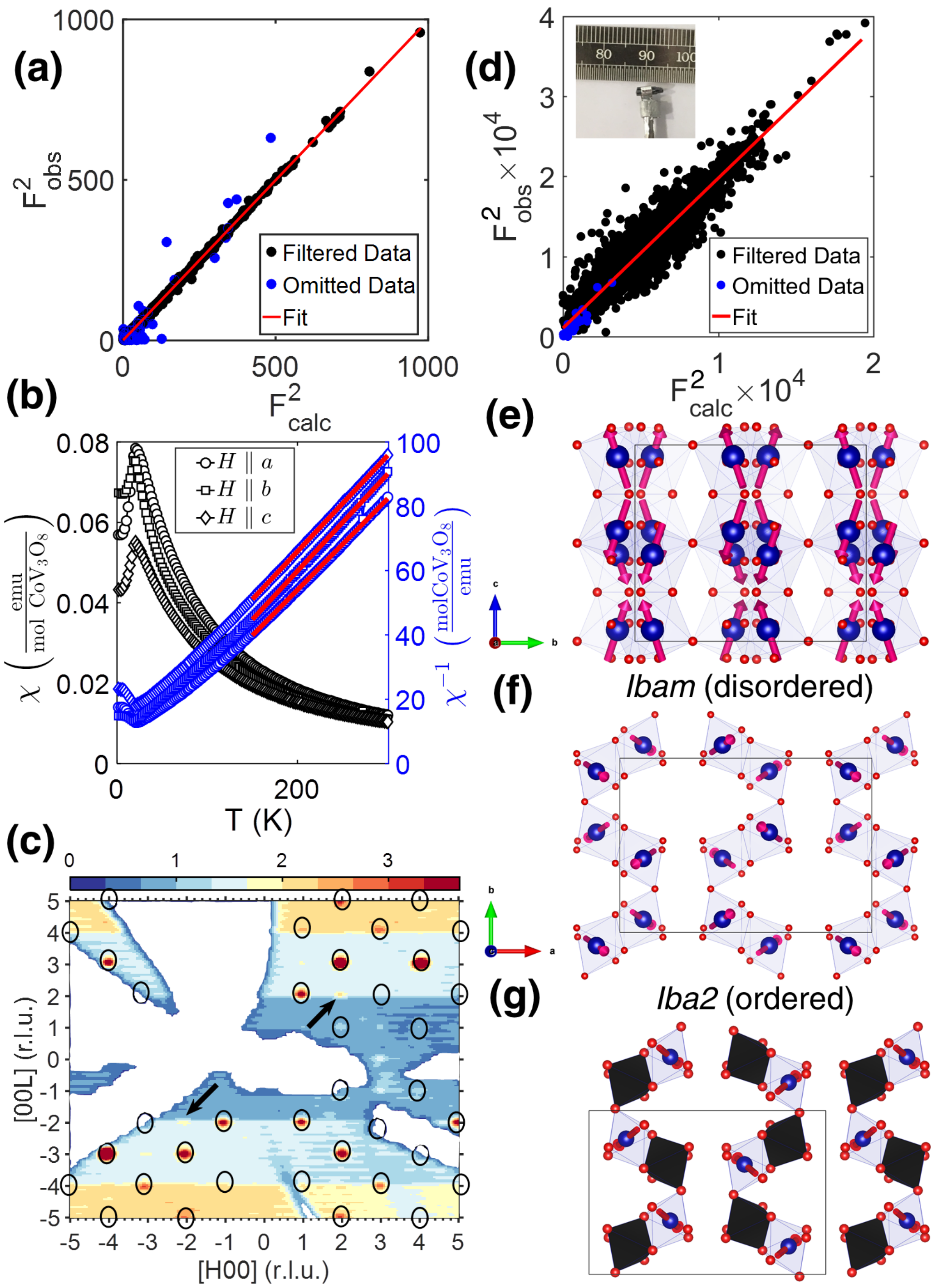}
	\caption{(a) Refinement of single crystal x-ray diffraction data collected at 120~K yielding a refined $Ibam$ unit cell ($a$=14.29344(4)~\AA, $b$=9.8740(3)~\AA, $c$= 8.34000(3)~\AA), in agreement with previous studies~\cite{oka}. (b) Temperature dependence of the DC magnetic susceptibility of \cvo~an external DC field $\rm{\mu_{o}H_{ext}}$~=~0.5~T applied parallel to the three principal axes. Red lines indicate Curie-Weiss fits to high temperature data and are summarized by Tab.~\ref{tab:CW}. (c) Single crystal neutron diffraction intensity pattern collected at 5 K in the ($H0L$) scattering plane. Black ellipses indicate nuclear Bragg reflections. Arrows indicate strong magnetic Bragg reflections at (-21-2) and (212). (d) Refinement of single crystal neutron diffraction data on a (inset) 0.4312~g single crystal of \cvo~collected at 5~K. Schematic illustration of the refined magnetic structure of \cvo~along the (e) $bc$ and (f) $ab$ planes with the Co$^{2+}$ having 50 \% occupancy. The orientation of the refined magnetic moments on Co$^{2+}$ are indicated by red arrows. (g) illustrates the ordered $Iba2$ space group with each Co$^{2+}$ site fully occupied and the black octahedra representing V$^{4+}$ sites.  Both panels (f) and (g) show a single layer of Co$^{2+}$ ions.}
	\label{fig:fig2}
\end{figure*}

\emph{Sample Preparation}: Single crystals of \cvo~were grown using a modified ``self-flux" heating routine for $\alpha$-CoV$_{2}$O$_{6}$~\cite{he}. Precursor polycrystalline samples of $\alpha$-CoV$_{2}$O$_{6}$ were first synthesized by a standard solid-state reaction consisting of heating a stoichiometric mixture of $\rm{Co(CH_{3}CO_{2})_{2}\cdot4~H_{2}O}$ (Sigma-Aldrich, 98\%) and V$_{2}$O$_{5}$ (Alfa Aesar, 99.6\%) in air for 12~h at 650$^{\circ}$C, then for 48~h at 725$^{\circ}$C, followed by quenching in liquid nitrogen~\cite{markkula,cov2o6}. A mixture of the $\alpha$-CoV$_{2}$O$_{6}$ polycrystalline precursor and V$_{2}$O$_{5}$ in a 3:2 ratio in the presence of approximately 0.01\% ($w/w$) of B$_{2}$O$_{3}$ (Alfa Aesar, 98.5\%) was heated in a vacuum sealed quartz tube at 780$^{\circ}$C for 24 h and subsequently cooled to 700$^{\circ}$C at a rate of 1$^{\circ}$C hr$^{-1}$. After 24~h of heating at 700$^{\circ}$C, the sample was cooled to 600$^{\circ}$C at a rate of 1$^{\circ}$C hr$^{-1}$ and subsequently quenched to room temperature.  \\ \\
\indent Polycrystalline samples of \cvo~and $\alpha$-ZnV$_{3}$O$_{8}$ were synthesized by a standard solid-state reaction consisting of heating a stoichiometric mixture of
CoO (Alfa Aesar, 95\%) or ZnO (Alfa Aesar, 99.99\%), VO$_{2}$ (Alfa Aesar, 99\%) and V$_{2}$O$_{5}$ for 96 h at 650$^{\circ}$C under static vacuum in a sealed quartz tube with intermittent grindings until laboratory powder x-ray diffraction confirmed no discernable impurities~\cite{ichikawa,znv3o8}. All stoichiometric mixtures of polycrystalline precursors were first mixed thoroughly together and finely ground to homogeneity with acetone using an agate mortar and pestle. The mixtures were pressed into $\sim$ 2~g pellets using a uniaxial press and subsequently placed in alumina crucibles or directly in quartz ampoules for reactions performed in air and in vacuum, respectively. Unless otherwise stated, all heating routines involved a ramping rate of 5$^{\circ}$C min$^{-1}$ and samples were furnace cooled back to room temperature. \\ \\
\indent \emph{Laboratory X-ray Diffraction}: Single crystal x-ray diffraction was performed at 120~K on a 0.011~mg single crystal of \cvo~with dimensions of 0.40~$\times$~0.11~$\times$~0.09~mm$^{3}$ using monochromated Mo K$_{\alpha}$ radiation on an Oxford Diffraction SuperNova dual wavelength diffractometer equipped with an Atlas CCD detector and an Oxford Cryostream-$Plus$ low-temperature device. Data collection, integration, scaling, multiscan absorption corrections and indexing were performed using the \texttt{CrysAlisPro v1.171.37.35e} software package~\cite{agilent2013agilent}. The structure solution was performed using a direct approach method with the \texttt{SHELXS-97} program in \texttt{Olex2}~\cite{olex2}. \\
\indent Room temperature powder diffraction patterns of $\alpha$-CoV$_{2}$O$_{6}$,~\cvo~and $\alpha$-ZnV$_{3}$O$_{8}$~were collected over 2$\theta$ = [5,~70]$^{\circ}$ in 0.0365$^{\circ}$ steps on a Bruker D2 Phaser laboratory x-ray diffractometer using monochromated Cu K$_{\alpha}$ radiation. \\ \indent All structural refinements for single crystal and polycrystalline measurements were performed using the \texttt{JANA2006}~\cite{JANA} and GSAS~\cite{GSAS} Rietveld refinement program packages, respectively, and are summarized in Appendix~\ref{sec:appendixA}. For the single-crystal refinement, the solved structure was refined by a full-matrix least squares against $F^{2}$ using only data $I>3\sigma(I)$. \\ \\
\indent \emph{DC Magnetic Susceptibility}: A 7.7~mg single crystal of \cvo~with dimensions of 2~$\times$~1~$\times$~1~mm$^{3}$ was aligned along the three principal axes. All crystal alignments were performed with polychromatic Laue backscattering diffraction employing adapted photostimulable plates using the Fujifilm FCR Capsula XL II system~\cite{laue}. The temperature dependence of ZFC magnetization for all three principal axes was measured on a Quantum Design MPMS in an external DC field $\rm{\mu_{o}H_{ext}}$~=~0.5~T applied parallel to the particular axis of interest. Measurements were performed in 2~K steps spaced linearly from 2~K to 300~K. \\ \\
\indent \emph{Neutron Single Crystal Diffraction}: Neutron single crystal diffraction experiments were performed on the SXD~\cite{SXD,SXD2} time-of-flight instrument at the ISIS spallation source. The SXD diffractometer employs the time-of-flight Laue technique. The combination of a polychromatic incident beam falling on a stationary sample surrounded by 11 ZnS scintillator PSDs covering $\Omega\sim2\pi$~sr enables quick access to a large amount of reciprocal space with minimal sample movement during data collection. A 0.4312~g single crystal of \cvo~with dimensions of 13.2~$\times$~4.1~$\times$~2.1~mm$^{3}$~as illustrated in Fig.~\ref{fig:fig2}(d) was mounted on the end of a 6 mm aluminum pin with aluminum tape, vertically suspended from a liquid helium 50 mm bore Orange cryostat providing $\omega$-motion in an accessible temperature range of 1.5 to 300~K. Diffraction data was collected at both 5~K and 50~K for three different single crystal frames with an accumulated charge of 1300 $\mu$A$\cdot$h ($\sim$ 8 h). After each temperature change, the sample was allowed to thermalize for 15 minutes. Reflection intensities were extracted, reduced and integrated to structure factors using standard SXD procedures, as implemented
in \texttt{SXD2001}~\cite{SXD3,SXD,SXD2}.  \\ \\
\indent \emph{Inelastic Neutron Time-of-Flight Scattering Spectroscopy}: All inelastic neutron scattering experiments were performed on the direct geometry MARI~\cite{MARI,MARI2} and indirect geometry IRIS~\cite{IRIS} time-of-flight spectrometers located at ISIS.  Neutron spectroscopic measurements were performed on powders as preliminary measurements found the signal from single crystals to be weak.  High-energy measurements ($>$ 2~meV) on 32.6~g of \cvo~and 31.9~g of $\alpha$-ZnV$_{3}$O$_{8}$ were performed on the direct geometry MARI spectrometer. The $\rm{t_{o}}$ chopper was operated at 50 Hz in parallel with a Gd chopper spun at 350, 300 and 250 Hz with incident energies E$\rm{_{i}}$~=~150,~60~and~15~meV, respectively, providing an elastic resolution of 5.87, 1.82 and 0.321~meV, respectively. A thick disk chopper spun at $f$~=~50~Hz reduced the background from high-energy neutrons. A top loading Displex CCR provided an accessible temperature range of 5 to 600~K. \\
\indent For lower energies, measurements on 15.1~g of \cvo~were performed on the indirect geometry IRIS spectrometer. As an indirect geometry spectrometer, the final energy E$\rm{_{f}}$ was fixed at 1.84 meV by cooled PG002 analyzer crystals in near backscattering geometry. The graphite analyzers were cooled to reduce thermal diffuse scattering~\cite{IRIS2}, providing an elastic resolution of 17.5~$\mu$eV. A top loading Displex CCR provided an accessible temperature range of 5 to 580~K. \\ \\
\indent \emph{Neutron Powder Diffraction}: Neutron diffraction measurements on 1.8~g of polycrystalline \cvo~were performed on the BT-4 thermal triple axis spectrometer~\cite{BT4} at the NIST Center for Neutron Research (NCNR). Incident and scattering neutron energies were set to 14.7~meV ($\lambda$=2.3592~\AA), selected by vertically focussing PG002 monochromator and analyzer crystals with PG filters located before and after the sample to reduce higher-order neutron contamination. The S\"{o}ller horizontal collimator configuration downstream order was 60$'$-monochromator-80$'$-sample-80$'$-analyzer-60$'$-detector. A top loading liquid helium 50 mm bore Orange cryostat provided an accessible temperature range of 1.5 to 300~K. $\theta$-2$\theta$ measurements were collected at both 3~K and 300~K over 2$\theta$~=~[15, 90]$\rm{^{\circ}}$ in 0.2$\rm{^{\circ}}$ steps ($|\mathbf{Q}|$~=~[0.695,3.766]~\AA$^{-1}$ in 0.009~\AA$^{-1}$ steps). Magnetic order parameter measurements were performed at 2$\theta$~=~41.6$\rm{^{\circ}}$ ($|\mathbf{Q}|$~=~1.89~\AA$^{-1}$) over T=[3, 32]~K in 0.1~K steps. 

 \section{Results \& Analysis} 

\subsection{Crystal Structure} ~\label{sec:1}

\indent As illustrated in Fig.~\ref{fig:fig2}(a) and summarized in Tabs.~\ref{tab:ap1}-\ref{tab:ap2} in Appendix~\ref{sec:appendixA}, single crystal x-ray diffraction at 120~K confirmed an orthorhombic unit cell ($a$~=~ 14.29344(4)~\AA, $b$~=~9.8740(3) \AA, $c$~=~8.34000(3)~\AA) with a volume of 1185.60(6)~\AA$^{3}$, corresponding to $Z$~=~8. Systematic extinctions provided $Ibam$ (\#72, Fig.~\ref{fig:1}(a,b)) and $Iba2$ (\#45, Fig.~\ref{fig:1}(d)) as possible space groups, with statistical analysis of the intensity data favoring the centrosymmetric $Ibam$. In a procedure analogous to previous studies on hydrothermally grown single crystals, the structure was solved using a direct method~\cite{oka}. The corresponding unit cell was found to consist of three metal sites with octahedral ($16k$), tetrahedral ($8j$) and trigonal bipyramidal ($8j$) coordination, with Co$^{2+}$ and V$^{4+}$ with half occupancies independently distributed over the 16$k$ site and V$^{5+}$ with full occupancies in the latter two $8j$ sites. Structural refinements utilizing 910 out of a total of 985 measured reflections confirmed two important conclusions from previous studies~\cite{oka,ichikawa}. Firstly, Co$^{2+}$ and V$^{4+}$ are both randomly and equally distributed over the $16k$ site with refined occupancies of 0.506(6) and 0.494(4), respectively. Secondly, the large refined anisotropic displacements resulting from placing the O(5) oxygen in the $8f$ position with full occupancy support the local selection rule consisting of Co$^{2+}$ and V$^{4+}$ occupying respective positions on opposite sides of the O(5) bridging oxygen ligand, as illustrated in Fig.~\ref{fig:1}(c).   

\subsection{DC Magnetic Susceptibility}

\indent As summarized by Tab.~\ref{tab:CW} and Fig.~\ref{fig:fig2}, DC susceptibility measurements along all three principal axes indicates that \cvo~behaves as a Curie-Weiss paramagnet at high temperatures and undergoes an antiferromagnetic transition at 19.5(5)~K, corresponding to a $T\rm{_{N}}$ much greater than previously reported T$\rm{_{N}}$ of 8.2~K for crystals grown hydrothermally~\cite{oka}. A fit of the high temperature data (Fig.~\ref{fig:fig2}$b$) to the Curie-Weiss law yielded Curie-Weiss temperatures $\theta\rm{_{CW}}$ of 9.5(7), 2(1), $-21.3(2)$~K for $\mu\rm{_{o}}H_{ext}$ applied along $a$, $b$ and $c$, respectively.  The small $\theta\rm{_{CW}}$ with an average $\theta\rm{_{CW}}$ = $-3.2(4)$~K is suggestive of either weak exchange interactions or the presence of multiple and nearly canceling ferro/antiferromagneic interactions resulting in the experimentally observed small average.  The differences in the constants measured along different directions is also indicative of an anisotropy in the system likely originating from the distortion of the local octahedra~\cite{cov2o6,buyers}. \\
\indent As illustrated in Fig.~\ref{fig:fig2}$(b)$, the magnetization does not approach zero in the low temperature limit after the antiferromagnetic transition. Instead, its value for all three principal axes plateaus at 2~K which indicates the possibility for the presence of paramagnetism at lower temperatures, although no measurements were conducted below 2~K. In contrast to the $d^{7}$ Co$^{2+}$ moments that can couple \emph{via} $e_{g}$ orbitals, coupling between the $d^{1}$ V$^{4+}$ moments are exclusively \emph{via} $t_{2g}$ orbitals which is predicted to be much weaker~\cite{rules1,rules2,rules3} and thus more likely to exhibit paramagnetic behavior. In fact, V$^{4+}$ paramagnetism is supported by the observation that the saturated moment in the low temperature limit corresponds to 0.150(2)~$\rm{\mu_{B}}$, a value consistent with a strongly reduced V$^{4+}$ effective paramagnetic moment predicted to occur in the presence of strong spin-orbit coupling and octahedral distortions as has been previously observed experimentally in Na$_{2}$V$_{3}$O$_{7}$~\cite{Ropka2004,gavilano}. Finally, the average effective paramagnetic moment of 5.213(7)~$\rm{\mu_{B}}$ is smaller than the predicted moment of 5.6~$\rm{\mu_{B}}$, for Co$^{2+}$ in an octahedral environment as studied in CoO~\cite{kanamori} and assuming a 1:1 ratio of high spin Co$^{2+}$ and V$^{4+}$, confirming that both spin-orbit and distortion effects play a significant role~\cite{markkula,kim12:85} in the magnetism of \cvo, a topic that will be later addressed with inelastic neutron scattering. 

 \begin{table}
	\caption{Curie-Weiss parameters\footnote{Calculated over a range of $150 \leq T \leq 300$~K} for \cvo~in an external DC field $\rm{\mu_{o}H_{ext}}$~=~0.5~T applied parallel to the three principal axes. Numbers in parentheses indicate statistical errors.}
	\begin{tabular}{ | c | c | c|c|} 
		\hline
		Crystallographic Axis & $C$ (emu K/mol) & $p\rm{_{eff}}$ ($\rm{\mu_{B}}$) & $\rm{\theta_{CW}}$ (K) \\ 
		\hline
		$a$ & 3.525(9) & 5.310(7) & 9.5(7) \\ 
		\hline
		$b$ & 3.31(2) & 5.15(2) & 2(1)\\ 
		\hline
		$c$	 & 3.354(2) & 5.180(2) & $-$21.3(2) \\ 
		\hline
		Average & 3.396(7) & 5.213(7) & $-$3.2(4) \\  
		\hline
	\end{tabular}
	\label{tab:CW}
\end{table}

\subsection{Magnetic Structure}  

\indent As shown in Fig.~\ref{fig:fig2}(c) and summarized in Tabs.~\ref{tab:ap3}-\ref{tab:ap4} in Appendix~\ref{sec:appendixA}, single crystal neutron diffraction confirmed both an absence of any structural distortion away from the $Ibam$ space group down to 5~K and the appearance of additional Bragg reflections confirming long range magnetic ordering as measured by previous DC susceptibility measurements~\cite{oka}. Since DC susceptibility measurements suggested that V$^{4+}$ remained paramagnetic down to at least 2~K, the refinement of single crystal neutron diffraction data collected at 5~K assumed that the magnetic Bragg reflections were exclusively due to Co$^{2+}$ that were randomly distributed throughout the 16$k$ metal sites. The random distribution of Co$^{2+}$ was accomplished by constraining the occupancy of each metal site to a value of $\frac{1}{2}$. The additional magnetic Bragg reflections were successfully indexed using a propagation vector $\mathbf{k}$~=~(1, 1, 1) with the $P_{I}ccn$ (\#56.376) Shubnikov space group~\cite{mag_group}. The propagation vector $\mathbf{k}$~=~(1,~1,~1) was initially chosen as it corresponds to the first point of symmetry reduction by removing body-centering symmetry with the same structural unit cell~\cite{degraef10:41_2}. Subsequently, utilizing the aforementioned value of $\mathbf{k}$, a symmetry analysis was performed in \texttt{JANA2006}~\cite{JANA}. 
With a $\mathbf{k}$~=~(1,~1,~1), the symmetry analysis considers which models were compatible --- following the symmetry operations of the group, but excluding body-centering --- with the restriction that moments at ($x$, $y$, $z$) are  antiferromagnetically aligned with those moments at ($x+\frac{1}{2}$, $y+\frac{1}{2}$, $z+\frac{1}{2}$). Four models were found to be compatible, with the $P_{I}ccn$ (\#56.376) Shubnikov space group yielding the best match. \\
\indent Tab.~\ref{tab:1} summarizes the results of a joint nuclear and magnetic refinement utilizing 5086 out of a total of 5120 measured reflections at 5~K~(Fig.~\ref{fig:fig2}(d)), confirming a strong preference for the $P_{I}ccn$ Shubnikov space group of $Ibam$ over $P_{I}cc2$ of $Iba2$.   The refined magnetic moment for Co$^{2+}$ was found to be $\mu$~=~3.53(2)~$\rm{\mu_{B}}$ with $\mu_{a}$, $\mu_{b}$ and $\mu_{c}$ as 1.35(5)~$\rm{\mu_{B}}$, $1.16(5)~\rm{\mu_{B}}$ and 3.05(5)~$\rm{\mu_{B}}$, respectively. \cvo~adopts a magnetic structure consisting of effective pairs of 2D layers in the $ab$ plane, separated from one another by a non-magnetic layer consisting of tetrahedrally coordinated V$^{5+}$, as illustrated in Fig.~\ref{fig:fig2}(f). Within these 2D layers, Co$^{2+}$ spins are ferromagnetically coupled along both $a$ and $b$, corresponding to inter-chain superexchange interactions. These 2D layers come in pairs with each offset from one another by [0.1858$a$, 0.1508$b$ and 0.1194$c$] with the pair being antiferromagnetically coupled to the adjacent pair along $c$, corresponding to intra-chain superexchange interactions.

\begin{table}
	\caption{Comparison of the refined magnetic moment's components assuming random ($Ibam$) and ordered ($Iba2$) distribution of Co$^{2+}$ and V$^{4+}$ on the metal sites of \cvo. The goodness-of-fit metric $\chi^{2}$ and residuals from the magnetic refinement of neutron single crystal diffraction data collected at 5~K suggests that Co$^{2+}$ and V$^{4+}$ are randomly distributed. Numbers in parentheses indicate statistical errors.}
	\begin{tabular}{ | c | c |c|} 
		\hline
		~~Parameter~~ & ~Value ($Ibam$)~ & ~Value ($Iba2$)~ \\ 
		\hline
		$\mu_{a}$& 1.35(5) $\rm{\mu_{B}}$ & 1.30(6) $\rm{\mu_{B}}$   \\ 
		\hline
		$\mu_{b}$	& 1.16(5) $\rm{\mu_{B}}$ & 1.09(8) $\rm{\mu_{B}}$ \\ 
		\hline
		$\mu_{c}$	& 3.05(4) $\rm{\mu_{B}}$ & 2.32(5) $\rm{\mu_{B}}$ \\  
		\hline
		$\chi^{2}$		& 3.18 &  5.15 \\ 
		\hline
		R$\rm{_{F^{2}}}$		& 8.38\% & 10.59\%   \\ 
		\hline
		R$\rm{_{wF^{2}}}$		& 8.99\% & 14.57\% \\
		\hline
		R$\rm{_{F^{2}_{mag}}}$		& 24.13\% & 32.28\%  \\ 
		\hline
	\end{tabular}
	\label{tab:1}
\end{table}

\subsection{Inelastic Neutron Scattering} 

\begin{figure*}
	\centering
	\includegraphics[width=1.0\linewidth]{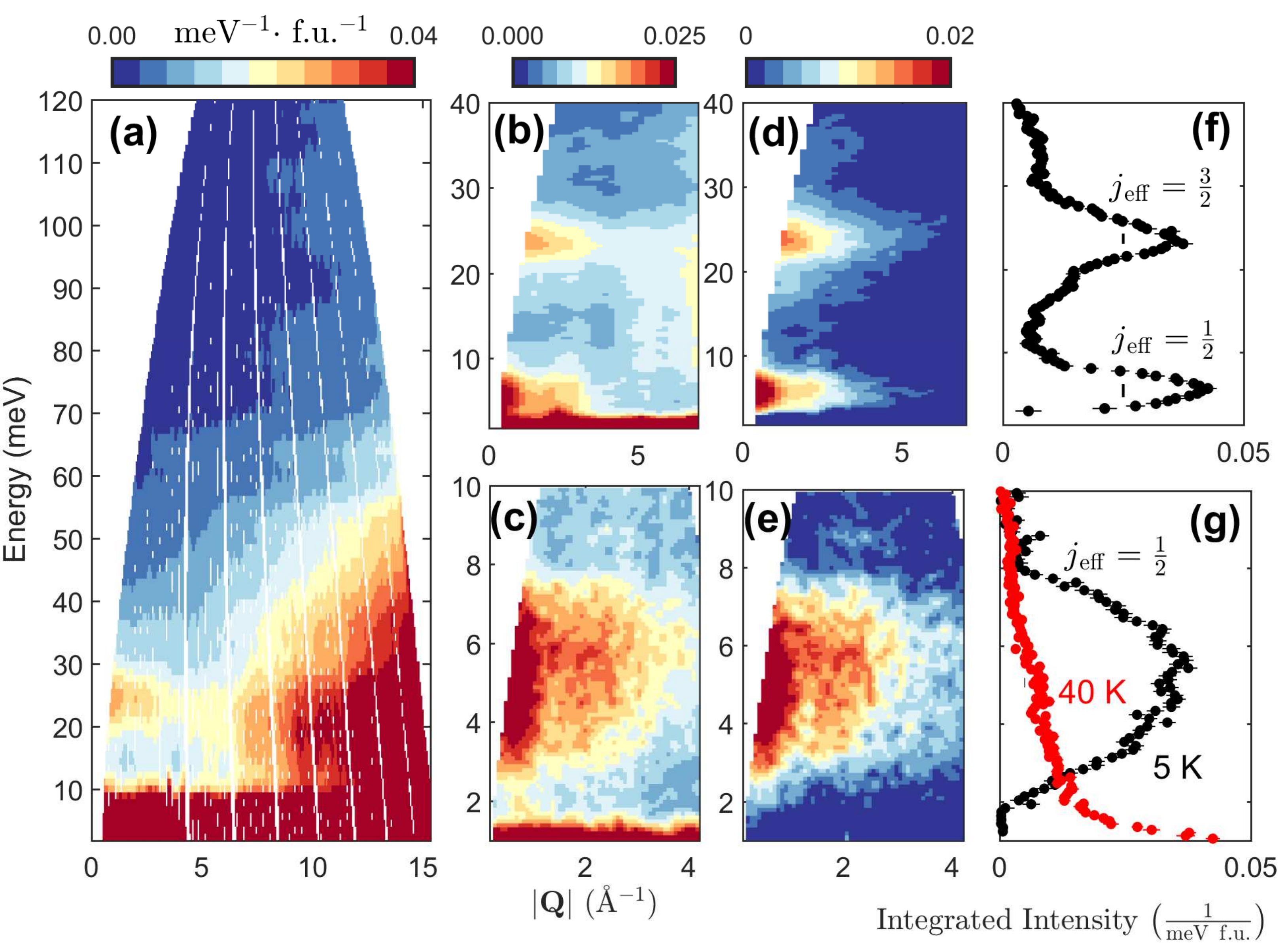}
	\caption{$\widetilde{S}(|\mathbf{Q}|,E)$ measured on MARI at T~=~5~K with an E$\rm{_{i}}$ of (a) 150~meV, (b) 60~meV and (c) 15~meV. (d,e) Magnetic scattering $\widetilde{S}_{M}(|\mathbf{Q}|,E)$ and (f,g) corresponding $|\mathbf{Q}|$-integrated cuts ($|\mathbf{Q}|$=[0,3]~\AA$^{-1}$). Vertical lines in (f,g) indicate instrumental resolution. $\widetilde{S}_{M}(|\mathbf{Q}|,E)$ was calculated by the subtraction of corresponding $\widetilde{S}(|\mathbf{Q}|,E)$ for  $\alpha$-ZnV$_{3}$O$_{8}$ measured at identical experimental conditions. All inelastic scattering intensities have been normalized to absolute units.}
	\label{fig:3}
\end{figure*}

\indent Motivated by the random distribution of Co$^{2+}$ and V$^{4+}$, multiple ferro-/antiferromagnetic interactions and the presence of strong spin-orbit coupling, the spin dynamics of \cvo~was investigated with inelastic neutron scattering. All inelastic scattering intensities were normalized to absolute units using the paramagnetic approximation~\cite{normalisation}. Normalization was performed by using both Co and V as internal incoherent standards~\cite{xu12:86,namno2} to calculate an absolute calibration constant $\mathcal{A}$ converting vanadium-corrected scattering intensities $\widetilde{I}(\mathbf{Q},E)$ to the differential scattering cross section $\frac{d^{2}\sigma}{dEd\Omega}$ which was then converted to the dynamic structure factor $S(\mathbf{Q},E)$ by  

\begin{equation}
\mathcal{A}\widetilde{I}(\mathbf{Q},E) \equiv \frac{d^{2} \sigma}{dEd\Omega} = \left(\frac{\gamma r_{o}}{2}\right)^{2}g^{2}_{J}2|f(\mathbf{Q})|^{2}S(\mathbf{Q},E),
\label{eq:definition}
\end{equation}    

\noindent where it is understood~\cite{fong00:61} that $S(\mathbf{Q},E)$ is $S^{zz}(\mathbf{Q},E)=\frac{Tr\left\{S^{\alpha\beta}(\mathbf{Q},E)\right\}}{3}$, $\left(\frac{\gamma r_{o}}{2}\right)^{2}$ and $g_{J}$ equals to 73~mb~sr$^{-1}$ and the Land\'{e} $g$-factor, respectively, while the factor of 2 corresponds to the paramagnetic cross section~\cite{squires_2012,namno2,normalisation,stock2004}. The value for the Land\'{e} $g$-factor is discussed in Appendix~\ref{sec:gj}.  Hereafter, all neutron scattering quantities with a tilde (for example $\widetilde{S}(|\mathbf{Q}|,E)$), denote the inclusion of the magnetic form factor squared $|f(\mathbf{Q})|^{2}$.

\subsubsection{Spin-Orbit Transitions} 

\indent As discussed in Appendix~\ref{sec:projection}, Co$^{2+}$ ($L=3$ and $S=\frac{3}{2}$) in an octahedral crystal field environment can be projected onto a ground state with an effective orbital angular momentum~\cite{buyers} of $l=1$ with a projection factor~\cite{cov2o6,kanamori} $\alpha$ of $-\frac{3}{2}$.   Diagonalizing the projected spin-orbit Hamiltonian $\hat{\mathcal{H}}_{SO}=\alpha\lambda \hat{\mathbf{l}}\cdot\hat{\mathbf{S}}$ results in three spin-orbit manifolds~\cite{cowley,buyers1968,projection} characterized by an effective angular momentum  $\hat{\mathbf{j}}=\hat{\mathbf{l}} + \hat{\mathbf{S}}$ with eigenvalues $j \equiv j\rm{_{eff}}$ of $\frac{1}{2}$, $\frac{3}{2}$, and $\frac{5}{2}$. The $j\rm{_{eff}}=\frac{3}{2}$ and $\frac{5}{2}$ manifolds are separated in energy from the $j\rm{_{eff}}=\frac{1}{2}$ ground state doublet manifold by $\frac{3}{2}\alpha\lambda$ and $\frac{5}{2}\alpha\lambda$, respectively~\cite{buyers}.  For pure CoO~\cite{cowley}, $|\alpha\lambda|\sim 24$~meV, and therefore for an undistorted octahedra, one would expect a crystal field excitation at $\sim$36~meV.  In this section, we study the magnetic excitations in \cvo~in order to determine if its ground state can be considered as a $j\rm{_{eff}}=\frac{1}{2}$. 

Given that only small single crystals were available of \cvo, preliminary neutron inelastic scattering data failed to produce a measurable signal.  To extract information on the low temperature magnetic dynamics, we therefore used powders and time-of-flight neutron spectroscopy techniques. Neutron inelastic scattering measurements (Figs.~\ref{fig:3}(a)-(c)) on polycrystalline \cvo~with an E$\rm{_{i}}$ = 150, 60 and 15~meV, respectively at 5~K revealed two clear low $|\mathbf{Q}|$ excitations at $\sim$ 5~meV and $\sim$ 25~meV. To prevent any weak magnetic signal of interest from being masked by strong phonon bands, a scaled inelastic scattering spectrum $\widetilde{\gamma}\widetilde{S}(|\mathbf{Q}|,E)$ of an approximate isostructural compound $\alpha$-ZnV$_{3}$O$_{8}$~\cite{znv3o8} collected with identical experimental conditions was subtracted as a background~\cite{sarte18:98}. Neutron inelastic scattering investigations of $\alpha$-ZnV$_{3}$O$_{8}$ on MARI found no evidence of correlated V$^{4+}$ moments over the energy range reported here.  The scaling factor $\widetilde{\gamma}$ for the background was calculated from the ratio between energy-integrated cuts of $\widetilde{S}(|\mathbf{Q}|,E)$  of \cvo~and $\alpha$-ZnV$_{3}$O$_{8}$ along $|\mathbf{Q}|$ at high $|\mathbf{Q}|$, thereby normalizing by the phonon scattering. The use of $\alpha$-ZnV$_{3}$O$_{8}$ as a background not only removes the constant and $|\mathbf{Q}|^{2}$-dependent background contributions but the presence of V$^{4+}$ in both compounds allows for the isolation of magnetic fluctuations solely attributable to Co$^{2+}$.  As illustrated in Figs.~\ref{fig:3}(d)-(g), the use of $\alpha$-ZnV$_{3}$O$_{8}$ as an effective background revealed that the origin of the low-$|\mathbf{Q}|$ excitations must be due to Co$^{2+}$ exclusively, excluding the possibility of any contribution from V$^{4+}$. 

Following the analysis of inelastic scattering measurements on monoclinic and triclinic polymorphs of CoV$_{2}$O$_{6}$~\cite{cov2o6}, the low-$|\mathbf{Q}|$ excitations in \cvo~can be understood as transitions between different spin-orbit manifolds. A comparison between the inelastic spectra of CoV$_{2}$O$_{6}$ and \cvo~suggests that the excitations at $\sim$ 5~meV and $\sim$ 25~meV are due to transitions within the $j\rm{_{eff}}=\frac{1}{2}$ manifold and between the $j\rm{_{eff}}=\frac{1}{2}$ and $j\rm{_{eff}}=\frac{3}{2}$ manifolds, respectively. In \cvo, these modes appear much broader than in CoV$_{2}$O$_{6}$; this will be discussed later. Such an assignment is supported by the observation that the transition at $\sim$ 5~meV is gapped at 5~K in the magnetically ordered regime, as illustrated in Figs.~\ref{fig:5}(a) and (b). Such a gap would be a consequence of the Zeeman splitting of the $j\rm{_{eff}}=\frac{1}{2}$ manifold due to the internal molecular field caused by long range ordering in the N\'{e}el phase~\cite{cov2o6}. Once the temperature is raised above T$\rm{_{N}}$, the molecular field would be significantly reduced due to the loss of magnetic order, resulting in the disappearance of a gap, as is experimentally observed in Fig.~\ref{fig:3}(g). 

In the context of this assignment in terms of $j\rm{_{eff}}$ spin-orbit split manifolds, a difference between \cvo~and monoclinic $\alpha$-CoV$_{2}$O$_{6}$ is the absence of an observable $\sim$110~meV magnetic excitation (Fig.~\ref{fig:3}(a)).  As was previously calculated for CoV$_{2}$O$_{6}$~\cite{cov2o6}, in addition to the strong excitations for the intra-$j\rm{_{eff}}=\frac{1}{2}$ and the $j\rm{_{eff}}=\frac{1}{2}$ to $j\rm{_{eff}}=\frac{3}{2}$ transitions, the intensity of the $j\rm{_{eff}}=\frac{1}{2}$ to $j\rm{_{eff}}=\frac{5}{2}$ transition scales with the distortion of the local coordination octahedra~\cite{cowley_lorentzian,martel} with the transition being absent for a perfect octahedra like in rocksalt and cubic CoO~\cite{cowley}.  The distortion of the local octahedra can quantified by the parameter $\delta$ defined by      

\begin{equation}
\delta = \frac{1}{\mathcal{N}}\sum\limits_{i} \left\{\left(\frac{d_{i} - \langle d \rangle}{\langle d \rangle}\right)^{2} \times 10^{4}\right\}, 
\label{eq:delta} 
\end{equation}

\noindent where $\mathcal{N}$~=~6 and $\langle d \rangle$ denotes the average distance~\cite{delta,cov2o6}. \cvo~exhibits a much weaker octahedral distortion ($\delta$~=~11.106(8)) than $\alpha$-CoV$_{2}$O$_{6}$ ($\delta$~=~55) and is thus expected to have a significantly weaker intensity.   This is also in agreement with previous results on triclinic $\gamma$-CoV$_{2}$O$_{6}$ (with $\delta$~=~2.1 and 4.8 for the two different Co$^{2+}$ sites) which failed to observe a $j\rm{_{eff}}=\frac{5}{2}$ transition~\cite{cov2o6}.   A distortion of the local octahedra around the Co$^{2+}$ site should result in an anisotropic term in the magnetic Hamiltonian~\cite{martel,gladney66:146,walter84:45}.  Given the powder average nature of the dynamics (discussed below), we are not sensitive to this term.   However, the consistency of the inelastic response with the $j\rm{_{eff}}$ description discussed above in terms of the energy response is also consistent with other Co$^{2+}$-based magnets where a local distortions of the octahedra exists~\cite{cowley_lorentzian,cov2o6}.  We now discuss further evidence for our interpretation in terms of $j\rm{_{eff}}$ levels by applying the sum rules of neutron scattering to the integrated inelastic scattering intensity.

\begin{figure}
	\centering
	\includegraphics[width=1.0\linewidth]{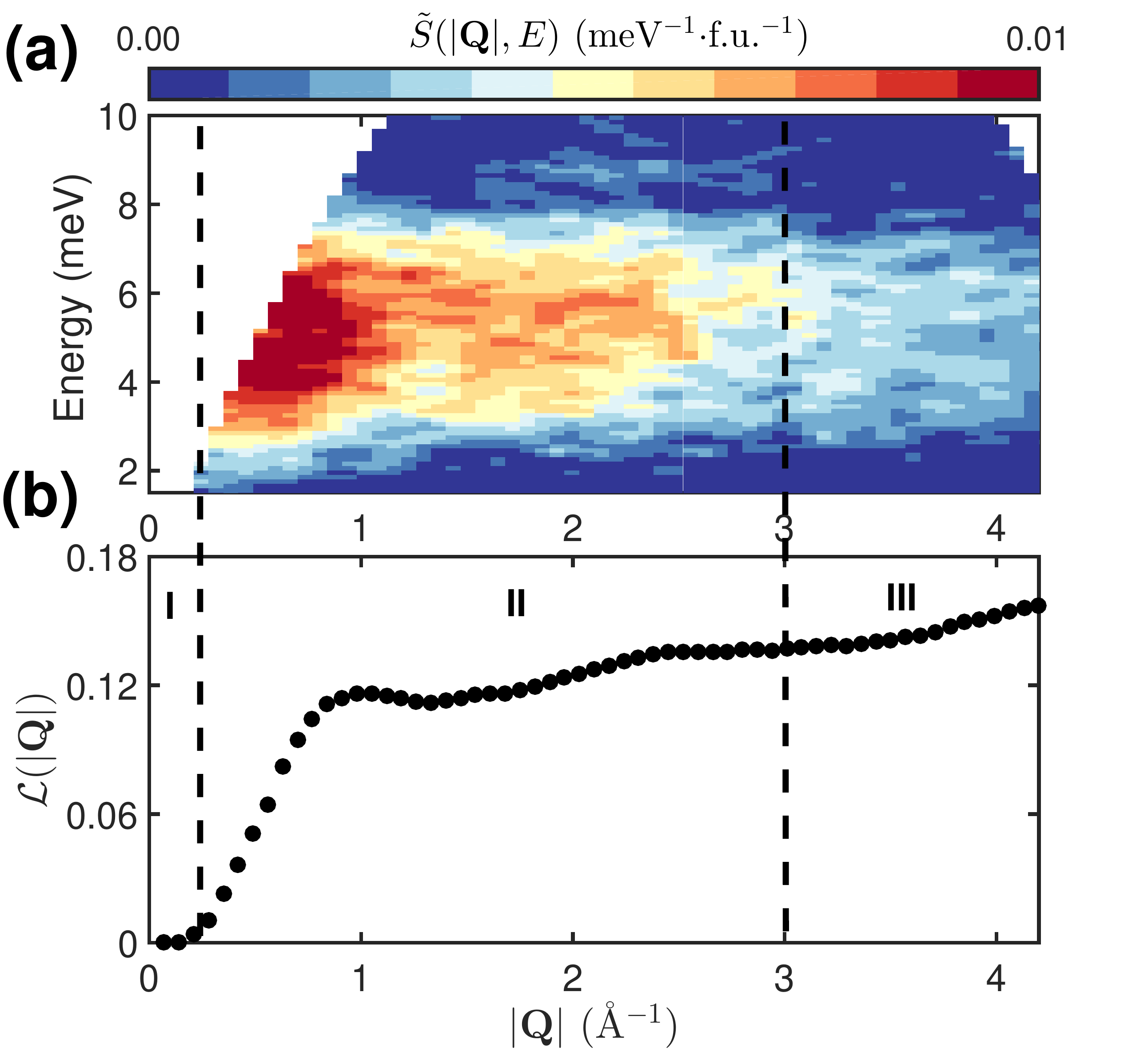}
	\caption{(a) Magnetic scattering $\widetilde{S}_{M}(|\mathbf{Q}|,E)$ of \cvo~measured on MARI at T~=~5~K with an E$\rm{_{i}}$ of 15~meV and the corresponding (b) $|\mathbf{Q}|$-dependence of the total integrated inelastic ($E$=[2,8]~meV) magnetic scattering intensity $\mathcal{L}$.  Regions I, II and III denote ``get-lost" tube-, magnetic- and phonon/form factor-dominated regions, respectively.}
	\label{fig:4}
\end{figure}

\subsubsection{Total Moment Sum Rule} \label{sec:total} 

To confirm the assignment of the 5~meV signal as excitations within the ground state $j\rm{_{eff}}=\frac{1}{2}$ manifold, the total integrated spectral weight at 5~K of the lowest lying excitation was calculated. As summarized by the total moment sum rule of neutron scattering~\cite{namno2,stone,sumrules,plumb,hammar98:57}, the sum of all spectral weight is defined by   
   
\begin{equation}
\frac{3\int d^{3}\mathbf{Q} \int dE S(\mathbf{Q},E)}{\int d^{3}\mathbf{Q}} = j(j+1),
\label{eq:totalmoment} 
\end{equation}

\noindent where $S(\mathbf{Q},E)\equiv S^{zz}(\mathbf{Q},E)$ denotes the magnetic component of the dynamic structure factor $\widetilde{S}^{zz}_{M}(\mathbf{Q},E)$ that has been further renormalized by $|f(\mathbf{Q})|^{2}$. The extra factor of 3 has been included to assure consistency with the definition of $S(\mathbf{Q},E) \equiv S^{zz}(\mathbf{Q},E)$ given by Eq.~\ref{eq:definition} in the paramagnetic approximation. A measurement of the integrated intensity is therefore sensitive to the effective $j$ of the manifold of levels being integrated over.  Eq.~\ref{eq:totalmoment} can be simplified by integrating out the angular dependence and canceling common terms resulting in an integral $\mathcal{L}$ defined by 

\begin{equation}
\mathcal{L}(|\mathbf{Q}|) = \frac{3\int d|\mathbf{Q} |\mathbf{Q}|^{2}\int dE S(|\mathbf{Q}|,E)}{\int d|\mathbf{Q}| |\mathbf{Q}|^{2}},
\label{eq:totalmoment2}
\end{equation}

\noindent The total integral $\mathcal{L}$ is uniquely a function of $|\mathbf{Q}|$ and represents an integration of the magnetic density of states over all energies including both elastic and inelastic channels in the cross section~\cite{namno2}.  With $j\rm{_{eff}}=\frac{1}{2}$, the total moment sum rule  (Eq.~\ref{eq:totalmoment}) would predict a value of 0.75 for the total integrated intensity. \\
\indent Since the assignment discussed above based on spin-orbit transitions assumes that the $\sim$5~meV excitation and the elastic cross section is exclusively due to excitations within the $j\rm{_{eff}}=\frac{1}{2}$ manifold, all quantities in Eq.~\ref{eq:totalmoment} were projected onto the ground state doublet manifold by the projection theorem of angular momentum~\cite{projection,abragam,sarte18:98}. As discussed in Appendix B2, the projection onto the ground state doublet required defining the projected value of the Land\'{e} $g$-factor $g_{J}$ as $g'_{J}=\frac{13}{3}$ and the effective angular momentum $j\rm{_{eff}}$ as $\frac{1}{2}$~\cite{KCoF3,buyers}.  As illustrated in Fig.~\ref{fig:4}, the total integrated inelastic intensity of $S(|\mathbf{Q}|,E) \equiv S^{zz}(|\mathbf{Q}|,E)$ given by $\mathcal{L}(|\mathbf{Q}|)$ (Eq.~\ref{eq:totalmoment2}) saturates at 0.15(1). Combining the total integral of the inelastic contribution and an elastic contribution~\cite{ybrh2si2} of $\left(\frac{\mu}{g'_{J}\mu_{B}}\right)^{2}=0.66$, yields a total integral of 0.81$\pm$0.14, in excellent agreement with the total moment prediction for $j\rm{_{eff}}=\frac{1}{2}$ of 0.75.  The agreement further confirms our assignment of the low energy excitations to transitions within the ground state $j\rm{_{eff}}=\frac{1}{2}$ spin-orbit doublet manifold.\\
\indent With the low energy excitations being successfully approximated by pure $j\rm{_{eff}}$ manifolds, we may now rationalize the effective paramagnetic moment $p\rm{_{eff}}$ of 5.213(7)~$\rm{\mu_{B}}$ that was calculated from DC susceptibility.  Given that the $j\rm{_{eff}}=\frac{1}{2}$ and $j\rm{_{eff}}=\frac{3}{2}$ manifolds are separated by $\sim$24~meV ($\sim$278~K), both are significantly thermally populated at the high temperatures used for the Curie-Weiss fit. In such a high temperature regime, we would expect a $p\rm{_{eff}}$ of $g_{s}\sqrt{S(S+1)}=3.9~\mu_{B}$, which is significantly less than the measured value as has been commonly observed for other magnets based on Co$^{2+}$ in octahedral coordination~\cite{cov2o6,kimber11:84,markkula,buyers1968}.  The extra component measured with susceptibility may be accounted for by noting that V$^{4+}$ contributes $g_{s}\sqrt{S(S+1)}$~=~1.7~$\rm{\mu_{B}}$.  Therefore the addition of the contributions to $p\rm{_{eff}}$ from both Co$^{2+}$ and V$^{4+}$ corresponds to a total predicted $p\rm{_{eff}}$~=~5.6~$\rm{\mu_{B}}$, in close agreement with the experimental data, with the small discrepancy potentially attributable to the fact that the $j\rm{_{eff}}=\frac{5}{2}$ manifold is still not significantly populated at T$\sim$300~K. Although it is worth noting that an additional and distinct possibility for a much larger measured effective paramagnetic moment may be a strong orbital contribution as has been observed for the case of CoO~\cite{radwanski04:345,neubeck01:62}, where the orbital contribution is significant, corresponding to approximately $\frac{1}{3}$ of the total ordered moment. 

\subsection{Critical Exponents}       

\indent Despite the similarities between the inelastic spectra of \cvo~and CoV$_{2}$O$_{6}$, one difference is the bandwidth of the low energy excitation that we have assigned to the $j\rm{_{eff}}=\frac{1}{2}$ manifold. As illustrated in Fig.~\ref{fig:3}(g), in contrast to both polymorphs of CoV$_{2}$O$_{6}$, \cvo~exhibits a broad peak in energy whose bandwidth is approximately 20 times that of instrumental resolution. Such a large bandwidth could be accounted for by magnetic exchange coupling between spins~\cite{cov2o6,exchange1,exchange2}. However, an alternative explanation may lie in the intrinsic cationic disorder inherent to the disordered $Ibam$ structure of \cvo~\cite{oka}. Such large cationic disorder would result in a distribution of cationic sites and correspondingly a spread of spin-orbit transitions as has been shown for multiple doped systems~\cite{songvilay,pyrochlore,pyrochlore2,pyrochlore3,rodriguez2007,inelastic,uemura}, and thus perhaps such disorder may also explain the large bandwidth in \cvo~due to a distribution of molecular fields splitting the $j\rm{_{eff}}=\frac{1}{2}$ manifold.  We investigate this possibility in this section using scaling.

\subsubsection{Scaling Analysis}

\begin{figure}
	\centering
	\includegraphics[width=1.0\linewidth]{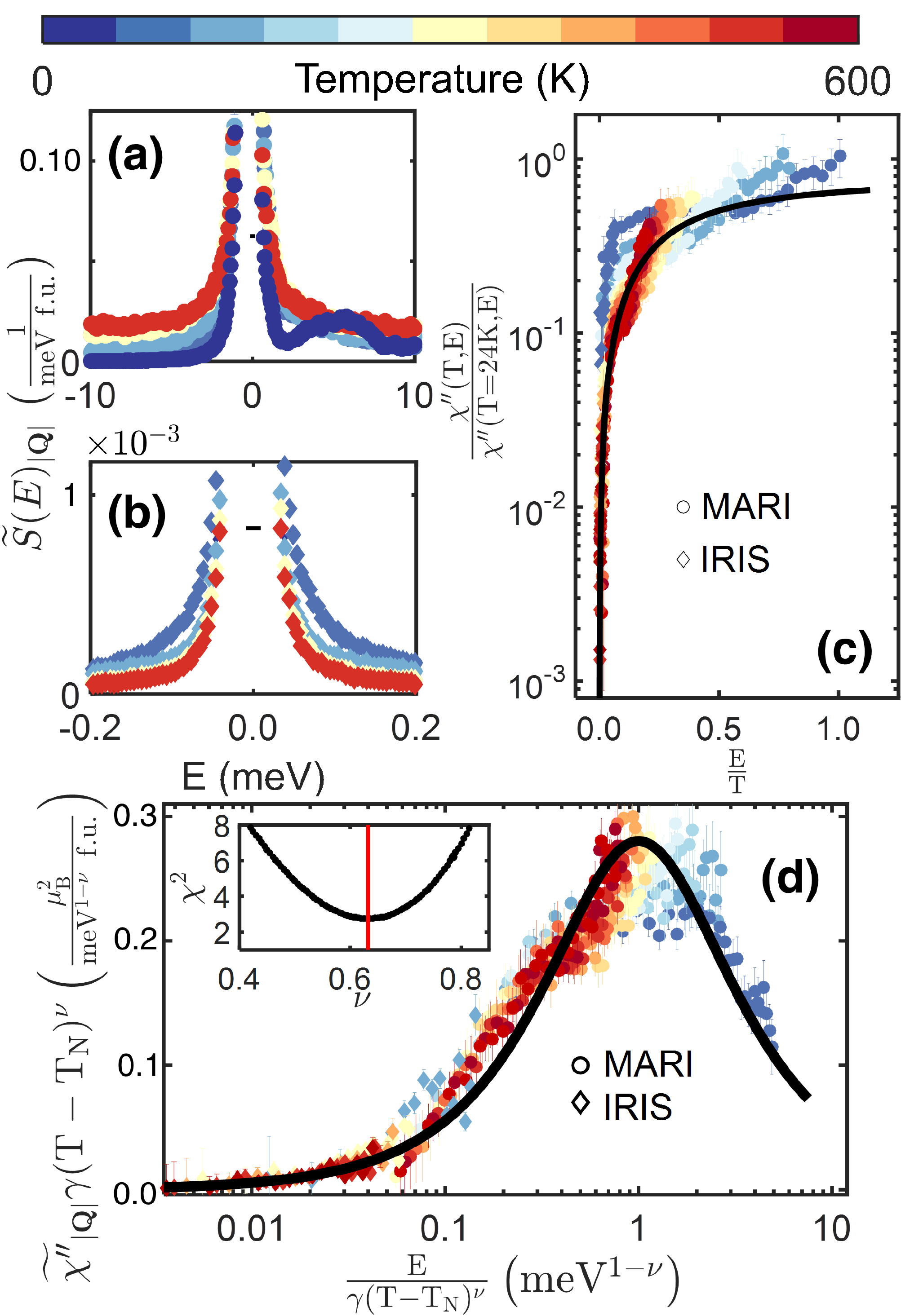}
	\caption{$|\mathbf{Q}|$-integrated cuts of $\widetilde{S}(|\mathbf{Q}|,E)$ measured on (a) MARI and (b) IRIS at various temperatures. Horizontal lines indicate instrumental resolution. (c) Energy and temperature dependence of the normalized $\chi''$ calculated from $|\mathbf{Q}|$-integrated cuts of $\widetilde{S}_{M}(|\mathbf{Q}|,E)$ measured on both IRIS at MARI. (d) Compilation of the energy-temperature dependence of $|\mathbf{Q}|$-integrated $\chi''$ as calculated in (c). As discussed in the main text, the data is described by a Lorentzian relaxational form (Eq.~\ref{eq:scaling}), revealing scaling behavior consistent with $\Gamma \propto (T-T{\rm{_{N}}})^{\nu}$. The line of best fit yields $\nu$~=~0.636(10), corresponding to a global minimum of $\chi^{2}$ as illustrated in the inset. All panels share the same temperature scale (top horizontal intensity bar). All $|\mathbf{Q}|$-integrated cuts on MARI and IRIS are from $|\mathbf{Q}|$=[0,3]~\AA$^{-1}$ for E$\rm{_{i}}$=15~meV and from $|\mathbf{Q}|$=[0,2]~\AA$^{-1}$ for  $E\rm{_{f}}$=1.84~meV, respectively.}
	\label{fig:5}
\end{figure}

The presence of such disorder would result in temperature being the dominant energy scale. To investigate this possibility, the temperature dependence of the Co$^{2+}$ spin fluctuations was analyzed using a scaling analysis previously employed for the charge doped cuprates~\cite{arctangent,arctangent2,bao2003,keimer,wilson2006}. For paramagnetic fluctuations, critical scattering theory assumes a single energy scale, the relaxation rate $\Gamma$, is dominant~\cite{collins}. If $\Gamma$ is driven by temperature, then it can be shown that the energy-temperature dependence of the uniform dynamic susceptibility $\chi''(E,T)$, follows $\frac{E}{T}$ scaling~\cite{arctangent,arctangent2} given by   
          
    \begin{equation}
    \frac{\chi''(T,E)}{\chi''(T=0~K,E)} = \arctan\left\{\sum\limits_{i=0}a_{i}\left(\frac{E}{T}\right)^{2i+1}\right\},
    \label{eq:arctan}
    \end{equation}
    
\noindent where $\chi''(T=0~K,E)$ denotes the value of $\chi''$ in the limit of T~=~0~K and all even powers are excluded in the sum to satisfy detailed balance, requiring $\chi''$ to be an odd function of energy~\cite{keimer}. For this particular analysis, the value of $\chi''(T,E)$ was calculated by first subtracting a temperature independent background from the measured $S(T,|\mathbf{Q}|,E)$. The contribution of the background was determined by an algorithm previously employed for Fe$_{1+x}$Te$_{0.7}$Se$_{0.3}$~\cite{background} and polymer quantum magnets~\cite{hong}. The algorithm is based on the requirement that all inelastic scattering must obey detailed balance accounting for both sample environment and other temperature-independent scattering contributions, and thus isolating the fluctuations exclusively due to Co$^{2+}$.  The background-subtracted dynamic structure factor was then converted to $\chi''(T,|\mathbf{Q}|,E)$ \emph{via} the fluctuation-dissipation theorem~\cite{mason}        

    \begin{equation}
    \chi''(T,|\mathbf{Q}|,E)~=~g^{2}\mu^{2}_{B}\pi\left\{\frac{1}{n(E,T)+1}\right\}S(T,|\mathbf{Q}|,E),  
    \end{equation}
    
\noindent where $n(E,T)$ is the Bose factor. Finally,  $\chi''(T,|\mathbf{Q}|,E)$ was integrated over $|\mathbf{Q}|$~=~[0, 3] and [0, 2]~\AA$^{-1}$ for measurements on MARI and IRIS, respectively. As illustrated in Fig.~\ref{fig:5}(c), $\frac{E}{T}$ scaling adequately accounts for the experimental data with the need for only two refined constants of 3.2(1) and 0.8(2) for $\rm{a_{o}}$ and $\rm{a_{2}}$, respectively, since the inclusion of higher order terms in Eq.~\ref{eq:arctan} did not improve the fit. The success of $\frac{E}{T}$ scaling suggests that $\Gamma \propto T^{\nu}$ and the larger value of $\rm{a_{o}}$ over all other terms suggests $\nu~\leq~1$.

The value of $\nu$ was refined using a modified scaling algorithm previously employed to detect anomalous scaling in the vicinity of a quantum critical point for CeCu$_{2}$Si$_{2}$ and CeCu$_{6-x}$Au$_{x}$~\cite{scaling3,scaling4,scaling}. Utilizing the single relational energy mode approximation and the Kramers-Kronig relations~\cite{jackson}, the uniform dynamic susceptibility can be approximated as a Lorentzian-like response~\cite{bean,cowley_lorentzian,hagen,arctangent,bao1998,Hutchings72:5,collins69:179,schullhof71:4} given by     

    \begin{equation}
    \chi'' = \chi'\left\{\frac{E\Gamma}{E^{2} + \Gamma^{2}}\right\},
    \label{eq:scaling0} 
    \end{equation}
    
\noindent where $\chi'$ denotes the static susceptibility and $\Gamma~\propto~\xi^{-1}$, where $\xi$ is the correlation length~\cite{chou91:43}. If one assumes both the single energy scale $\Gamma~= ~\gamma\rm{(T-T_{N})^{\nu}}$ and the static susceptibility $\chi' = \frac{C}{\Gamma}$, where $\gamma$ and $C$ are constants, then Eq.~\ref{eq:scaling0} assumes the form    

\begin{equation}
    \chi'' = \frac{C}{\gamma(T-T_{N})^{\nu}}\left\{\frac{\frac{E}{\gamma(T-T_{N})^{\nu}}}{1+\left(\frac{E}{\gamma(T-T_{N})^{\nu}}\right)^{2}} \right\}. 
    \label{eq:scaling} 
 \end{equation}

\noindent The first assumption leading to Eq.~\ref{eq:scaling} stems from the fact that the scaling properties of the dynamics are being investigated near the vicinity of an ordering transition at $\rm{T_{N}} \sim 19$~K and not a quantum critical point as in the cuprates and heavy fermion systems~\cite{ybrh2si2,scaling,scaling2}, a fact that was reflected in Fig.~\ref{fig:5}(c) by defining $\chi''(T=0,E)$ as the value at 24~K. The second assumption is based on the paramagnetic behavior observed with DC susceptibility at high temperatures, suggesting $\chi'$ should adopt a Curie-Weiss form~\cite{ybrh2si2,arctangent}. As illustrated in Fig.~\ref{fig:5}(d), the scaling relation (Eq.~\ref{eq:scaling}) provides a good description of the experimental data over 4 orders of magnitude in $\frac{E}{T}$, yielding a refined $\nu$ of 0.636(10).  It is important to note that the refined value of $\nu$ is not consistent with random dilute 3D Ising behavior where $\nu=$~0.683(2), but instead is consistent with the ordered 3D Ising universality class with a $\nu=$~0.6312(3)~\cite{universality,ballesteros1998,calabrese2003,calabrese2003b,collins,george}.  While scaling and critical scattering typically only applies near the phase transition, work on other transition metal based compounds has found critical scattering that scales up to high temperatures in the paramagnetic regime~\cite{stock11:83,exchange1}.

\begin{figure}
	\centering
	\includegraphics[width=1.0\linewidth]{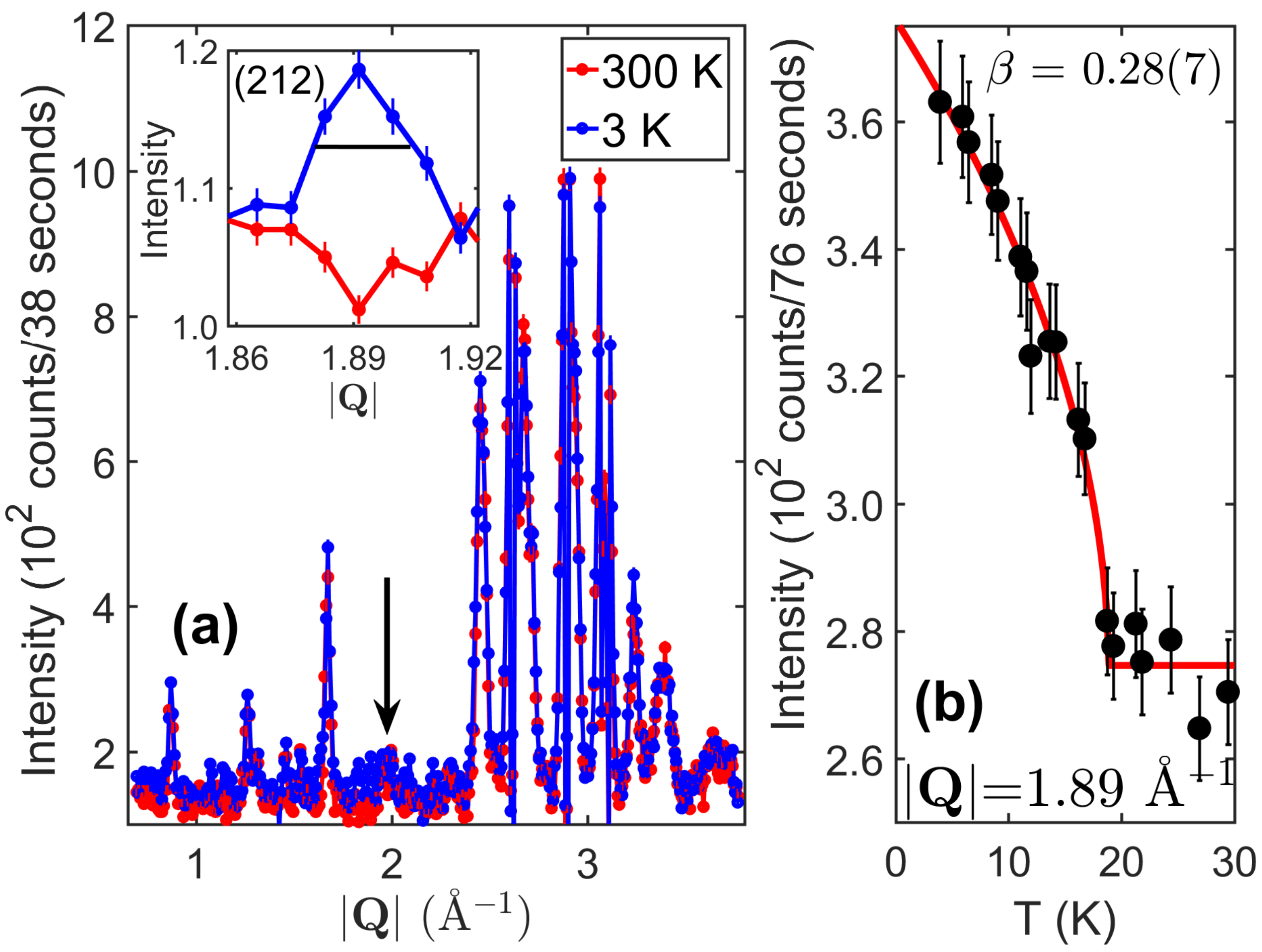}
	\caption{(a) Neutron diffraction profiles of polycrystalline \cvo~collected at 3 and 300 K on BT4. (inset) Additional scattering intensity on the (212) magnetic Bragg reflection at 3~K confirms long range magnetic order. The horizontal line indicates instrumental resolution. (b) Temperature dependence of the elastic intensity at $|\mathbf{Q}|$~=~1.89~\AA$^{-1}$ (2$\theta$~=~41.6$^{\circ}$), corresponding to the maximum of the (212) magnetic Bragg reflection as indicated by the arrow in (a). A fit to $\rm{(T_{N}-T)^{2\beta}}$ yields T$\rm{_{N}}$=18.8(6)~K and $\beta$=0.28(7).}
	\label{fig:8}
\end{figure}

\subsubsection{Magnetic Order Parameter} 

\indent The scaling analysis in the previous section found that the critical fluctuations are both consistent with an ordered three dimensional Ising universality class and with the DC susceptibility data presented above.  Consequently, while the excitations are separated into distinct $j\rm{_{eff}}$ manifolds, the scaling analysis indicates that the distortion does introduce an anisotropy term in the magnetic Hamiltonian influencing the critical dynamics outlined in the previous section. In an attempt to further deduce the universality class of CoV$_{3}$O$_{8}$, neutron diffraction measurements were performed on polycrystalline \cvo~to extract further critical exponents. As illustrated in Fig.~\ref{fig:8}(a), polycrystalline \cvo~exhibits long range magnetic ordering at 3~K, in agreement with both single crystal DC susceptibility (Fig.~\ref{fig:fig2}(b)) and single crystal neutron diffraction (Fig.~\ref{fig:fig2}(c)) measurements. The temperature dependence of the scattering intensity of the (212) magnetic Bragg reflection is displayed in Fig.~\ref{fig:8}(b), corresponding to the square of the magnetic order parameter~\cite{orderparameter} $\phi$, given by the power-law dependence               

\begin{equation}
I(T) \equiv \phi^{2}(T) \propto (T{\rm{_{N}}}-T)^{2\beta},
\label{eq:order_parameter}
\end{equation}

\noindent yields a refined $\rm{T_{N}}$ of 18.8(6)~K in agreement with DC susceptibility measurements and a refined $\beta$ of 0.28(7). Although the value of $\beta$ is in agreement with the predicted value of 0.326 for the ordered 3D Ising universality class~\cite{collins}, the large statistical error also implies agreement with the predicted value for the random dilute 3D Ising model of 0.35~\cite{universality,ballesteros1998,calabrese2003,calabrese2003b}. Therefore, the critical magnetic fluctuations are in agreement with expectations from both ordered and disordered 3D Ising behavior.    

\subsection{First Moment Sum Rule, Local Cation Ordering \& Single Mode Approximation}

\begin{figure*}[!htb]
	\centering
	\includegraphics[width=1.0\linewidth]{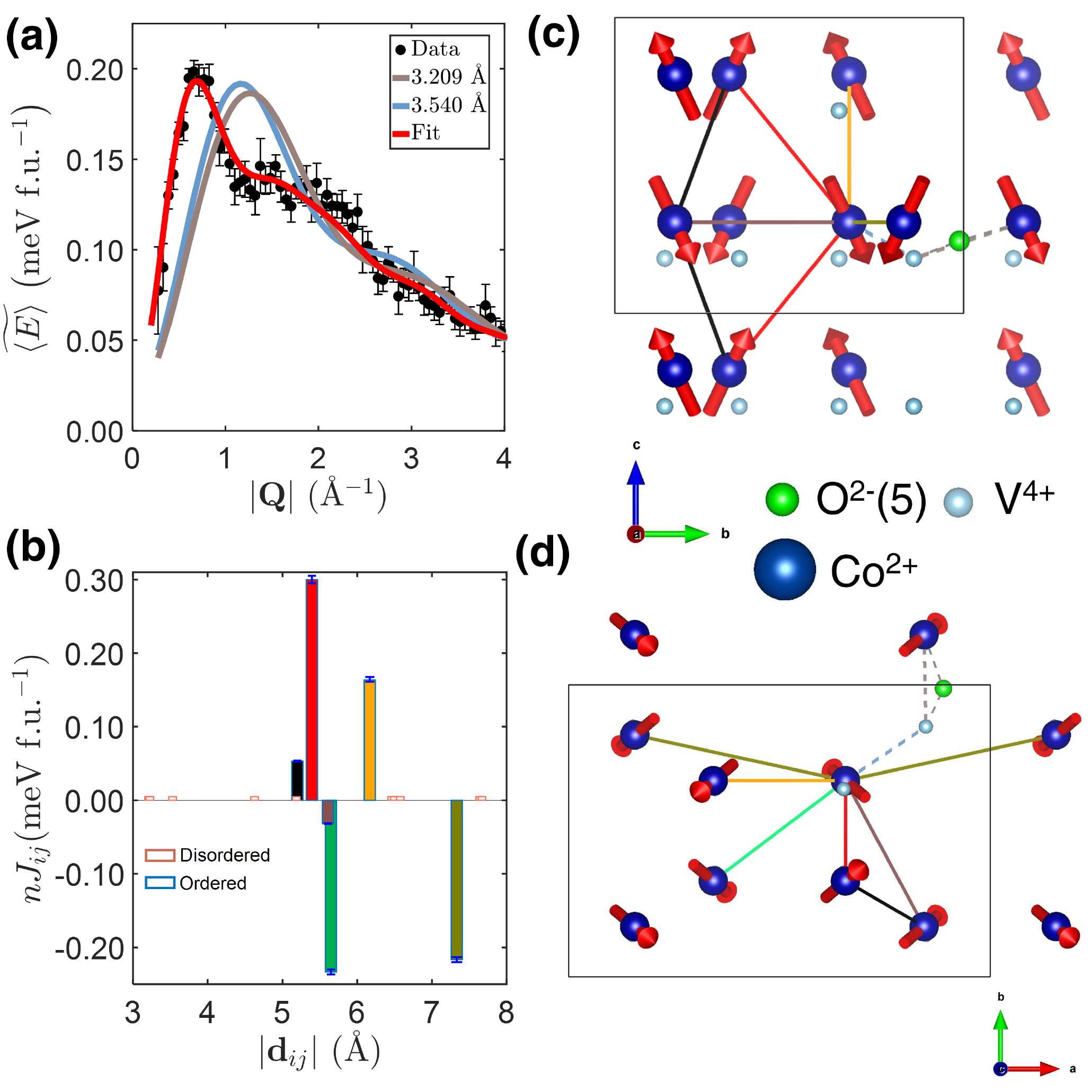}
	\caption{(a) $|\mathbf{Q}|$-dependence of the background subtracted first moment $\widetilde{\langle E \rangle}$ as measured on MARI at T=5~K with an E$\rm{_{i}}$=15~meV integrated over E=[2,8]~meV. A fit to the first moment sum rule (Eq.~\ref{eq:firstmoment2}) reveals that only six distances $|\mathbf{d}_{ij}|$ out to 7.5~\AA~possess non-negligible $nJ_{ij}$ values as illustrated in (b), and summarized in Tab.~\ref{tab:2}. For the purposes of comparison, distances present only in the ordered and disordered atomic arrangements are distinguished by purple and dark pink outline colors, respectively. Distances with non-negligible $nJ_{ij}$ contributions have a face color corresponding to the illustration of the corresponding six interactions along the (c) $bc$ and (d) $ac$ planes of the \cvo~unit cell. Both non-bridging oxygen atoms have been excluded and V$^{4+}$ ions have been reduced in size for the purposes of clarity. Two particular distances: 3.209~\AA~and 3.540~\AA~are absent as noted in (a), corresponding to nearest neighbor and bridging metal site distances, respectively. }
	\label{fig:6}
\end{figure*}

\indent In order to deduce further information concerning both the dimensionality $d$ and the microscopic exchange constants $J$, a combination of the first moment sum rule of neutron scattering and the single mode approximation was employed.  The determination of the values for $J$ and $d$ begin with the Hohenberg-Brinkman first moment sum rule~\cite{hohenberg} given by

\begin{align}
\langle E \rangle(\mathbf{Q}) &= \int dE E S(\mathbf{Q},E) \nonumber \\
&= -\frac{2}{3}\sum\limits_{i,j}n_{ij}J_{ij}\langle \hat{\mathbf{S}}_{i} \cdot \hat{\mathbf{S}}_{j} \rangle(1-\cos(\mathbf{Q} \cdot \mathbf{d}_{ij}))
\label{eq:firstmoment}
\end{align}

\noindent and its powder-average 

\begin{equation}
\langle E \rangle(|\mathbf{Q}|) = -\frac{2}{3}\sum\limits_{i,j} n_{ij}J_{ij}\langle\hat{\mathbf{S}}_{i}\cdot\hat{\mathbf{S}}_{j}\rangle\left\{1-\frac{\sin(|\mathbf{Q}||\mathbf{d}_{ij}|)}{|\mathbf{Q}||\mathbf{d}_{ij}|} \right\},
\label{eq:firstmoment2} 
\end{equation}

\noindent as derived in Appendix C, where $n_{ij}$, $J_{ij}$, $\langle \hat{\mathbf{S}}_{i} \cdot \hat{\mathbf{S}}_{j} \rangle$ and $\mathbf{d}_{ij}$ denote the number of individual exchange interactions, the exchange constant, the spin-spin correlator and the displacement vector between spins at sites ${i}$ and ${j}$, respectively~\cite{plumb,dimer,cov2o6}. \\
\indent Since all of the inelastic intensity measured at 5~K on MARI with an $E\rm{_{i}}$~=~15~meV shown in Fig.~\ref{fig:3}(g) corresponds to excitations within the ground state $j\rm{_{eff}}=\frac{1}{2}$, proven by the total moment sum rule, then the single mode approximation (SMA) can be applied~\cite{xu,plumb}. The single mode approximation, applicable to a situation where the excitation spectrum is dominated a single coherent mode, allows for the dynamic structure factor to be written as $S(\mathbf{Q},E)=S(\mathbf{Q})\delta[\epsilon(\mathbf{Q})-E]$,  where $\delta[\epsilon(\mathbf{Q})-E]$ assures energy conservation~\cite{namno2,cov2o6,stone,hong,mccabe2014}. Applying the single mode approximation to the first moment sum rule yields

\begin{align}
S(\mathbf{Q},E)  &=  -\frac{2}{3}\frac{1}{\epsilon(\mathbf{Q})}\sum\limits_{i,j}n_{ij}J_{ij}\langle \hat{\mathbf{S}}_{i} \cdot \hat{\mathbf{S}}_{j} \rangle \cdot \nonumber\\
&\qquad {} \left\{1 -\cos(\mathbf{Q}\cdot \mathbf{d}_{ij}) \right\}\delta[\epsilon(\mathbf{Q})-E],
\label{eq:singlemodeapprox}
\end{align}

\noindent providing a quantitative relationship between $S(\mathbf{Q},E)$ and the dispersion $\epsilon(\mathbf{Q})$ and by extension, a measure of the dimensionality~\cite{cov2o6,namno2,SMA,SMA2}. For numerical purposes, the delta function was approximated as a Lorentzian with a FWHM equal to that of the calculated experimental resolution width of 0.24~meV at 5~meV transfer on MARI. Eqs.~\ref{eq:firstmoment}-\ref{eq:singlemodeapprox} assume the presence of Heisenberg exchange and thus excludes exchange anisotropy~\cite{stone,xu,cov2o6,plumb}. It is important to note that the exclusion of any anisotropy terms is simply a first approximation based on the success of the isotropic exchange model to account for experimental data in a variety of other Co$^{2+}$-based systems such as CoV$_{2}$O$_{6}$, KMn$_{1-x}$Co$_{x}$F$_{3}$ and Mn$_{1-x}$Co$_{x}$F$_{2}$~\cite{cov2o6,cowley1972properties,KCoF3,svensson,buyers1968}. In fact, there is evidence that anisotropic exchange is not negligible in \cvo. Such experimental evidence includes equal intensities for transitions within the ground state manifold and between the ground state and first excited state manifolds~\cite{buyers1968,buyers}, as illustrated in Fig.~\ref{fig:3}(f). Another piece of evidence is the presence of a weak signal at $\sim$~1~meV at low energy transfer measurements, as illustrated in Fig.~\ref{fig:3}(g) that may be indicative of anisotropic breakdown of magnetic excitations~\cite{cov2o6,stone,namno2,buyers,ramazanoglu2009}. The non-negligible value of anisotropic exchange in \cvo~is indeed expected due to the distorted octahedra around Co$^{2+}$ ($\delta \sim$ 11) and has been observed in $\alpha,\gamma$-CoV$_{2}$O$_{6}$ with similar distortion parameters~\cite{cov2o6,kim12:85} but will be excluded in the context of the current discussion. 

\subsubsection{First Moment Sum Rule \& Cation Order} 

\indent   This section utilizes the first moment sum rule of neutron scattering to provide an estimate of the exchange constants in \cvo. Fig.~\ref{fig:6}(a) shows the background subtracted first moment $\widetilde{\langle E \rangle}(|\mathbf{Q}|)$ at 5~K was successfully described by the powder averaged first moment sum rule (Eq.~\ref{eq:firstmoment2}) incorporating all possible 15 Co$^{2+}-$Co$^{2+}$ distances in the \cvo~unit cell from $|\mathbf{d}_{ij}|$~=~[3.209, 7.669]~\AA. As summarized by Tab.~\ref{tab:2}, a least squares optimization revealed that only six unique distances possess non-negligible $-n_{ij}J_{ij}\langle \hat{\mathbf{j}}_{i} \cdot \hat{\mathbf{j}}_{j}\rangle$ values, where the use of $\hat{\mathbf{j}}$ in the correlator instead of $\hat{\mathbf{S}}$ is due to the use of $g'_{J}$ in the normalization process. Two particular distances with negligible $-n_{ij}J_{ij}\langle \hat{\mathbf{j}}_{i} \cdot \hat{\mathbf{j}}_{j}\rangle$ contributions are 3.209~\AA~and 3.540~\AA, corresponding to the nearest neighbor and metal site distances across the O(5) bridging ligand, respectively. The absence of the latter is expected due to the local selection rule~\cite{oka} as illustrated in Fig.~\ref{fig:1}(c), but the absence of the nearest neighbor distance is inconsistent with a random distribution of Co$^{2+}$ inherent to the disordered $Ibam$ structure previously deduced by diffraction measurements that are summarized in Fig.~\ref{fig:fig2}. Upon closer inspection of the \cvo~unit cell, these six distances were shown to correspond to the unique distances found exclusively in the ordered $Iba2$ structure~\cite{znv3o8} as illustrated in Figs.~\ref{fig:6}(c) and (d), confirming an ordered arrangement of Co$^{2+}$. \\
\indent While this analysis indicates the distances are consistent with the ordered $Iba2$ structure, there are two potential caveats. Because we measure the product $-n_{ij}J_{ij}\langle \hat{\mathbf{j}}_{i} \cdot \hat{\mathbf{j}}_{j}\rangle$, (i) the value of $n_{ij}$ may not be negligible but instead it may be the correlator $\langle \hat{\mathbf{j}}_{i} \cdot \hat{\mathbf{j}}_{j}\rangle$ whose value is negligible; (ii) and/or the exchange constants $J_{ij}$ may themselves be negligible. 
To address issue (i), we have calculated the correlator $\langle \hat{\mathbf{j}}_{i} \cdot \hat{\mathbf{j}}_{j} \rangle$ based on energy-integrated magnetic diffraction data (Tab.~\ref{tab:2}) and found it to be substantial for all distances.  We address argument (ii) by pointing out that some distances with negligible $-n_{ij}J_{ij}\langle \hat{\mathbf{j}}_{i} \cdot \hat{\mathbf{j}}_{j}\rangle$ have a Co$^{2+}$-O$^{2-}$-Co$^{2+}$ angle close to 180$^{\circ}$, predicted by the Goodenough-Kanamori-Anderson rules to yield strong antiferromagnetic exchange~\cite{rules1,rules2,rules3}. \\ 
\begin{table}[!]
	\caption{Distances $|\mathbf{d}_{ij}|$ with corresponding non-negligible refined values of $-n_{ij}J_{ij}\langle \hat{\mathbf{j}}_{i} \cdot \hat{\mathbf{j}}_{j}\rangle$ and $n_{ij}J_{ij}$ from the fit of the first moment $\langle E \rangle (|\mathbf{Q}|)$ (E=[2,8]~meV) at 5~K to the first moment sum rule~\cite{hohenberg}. The corresponding calculated spin-orbit corrected Curie-Weiss constant $\widetilde{\theta}_{CW}$ (Eq.~\ref{eq:Curie-Weiss}) is in close agreement with the experimentally determined Curie-Weiss constant averaged over all three principal directions $\bar{\theta}\rm{_{CW,exp}}$. Numbers in parentheses indicate statistical errors.}
	\renewcommand{\arraystretch}{1.3}
	\begin{tabular}{ | c | p{2.1cm}| c|c|} 
		\hline
		$|\mathbf{d}_{ij}|$ (\AA) & $-n_{ij}J_{ij}\langle \hat{\mathbf{j}}_{i} \cdot\hat{\mathbf{j}}_{j}\rangle$ (meV f.u.$^{-1}$) &~~$\langle \hat{\mathbf{j}}_{i} \cdot \hat{\mathbf{j}}_{j}\rangle$~~& $n_{ij}J_{ij}$ (meV f.u.$^{-1}$) \\ 
		\hline
		\textcolor{model1}{5.200(2)} & ~~~~~\textcolor{model1}{0.023(1)} &
		\textcolor{model1}{$-$0.420(2)}&
		\textcolor{model1}{0.055(1)} \\ 
		\hline
		\textcolor{model2}{5.395(3)} & ~~~~~\textcolor{model2}{0.173(1)} &
		\textcolor{model2}{$-$0.594(3)}& 
		\textcolor{model2}{0.30(1)} \\ 
		\hline
		\textcolor{model3}{5.6083(14)} & ~~~~~\textcolor{model3}{0.016(2)} & 
		\textcolor{model3}{0.484(2)} &
		\textcolor{model3}{$-$0.033(1)}\\ 
		\hline
		\textcolor{model4}{5.649(4)}  & ~~~~~\textcolor{model4}{0.099(2)} & 
		\textcolor{model4}{0.417(3)} &
		\textcolor{model4}{$-$0.24(1)}\\ 
		\hline
		\textcolor{model5}{6.168(3)} & ~~~~~\textcolor{model5}{0.08(1)} & 
		\textcolor{model5}{$-$0.483(3)} &
		\textcolor{model5}{0.17(1)}\\ 
		\hline
		\textcolor{model6}{7.3321(9)}  & ~~~~~\textcolor{model6}{0.13(1)} & 
		\textcolor{model6}{0.595(4)} &
		\textcolor{model6}{$-$0.22(1)}\\
		\hline
		$\bar{\theta}\rm{_{CW,exp}}$ & \multicolumn{3}{c|}{$-$3.2(4)~K} \\ 
		\hline
		$\widetilde{\theta}\rm{_{CW}}$ & \multicolumn{3}{c|}{$-$0.24(15)~K}  \\	
		\hline
	\end{tabular}
	\label{tab:2}
\end{table} 
\indent We now extract the exchange constants $J_{ij}$ by dividing out the correlator from the $-n_{ij}J_{ij}\langle \hat{\mathbf{j}}_{i} \cdot \hat{\mathbf{j}}_{j}\rangle$. By inserting the 6 values of $n_{ij}J_{ij}$ in the mean field expression for the Curie-Weiss temperature~\cite{meanfield,kittel} 

\begin{equation}
\widetilde{\theta}_{\rm{CW}} = -\frac{S(S+1)\sum\limits_{i,j}n_{ij}J_{ij}}{3\zeta},
\label{eq:Curie-Weiss}
\end{equation}

\noindent where $\zeta$ is a scale factor of 1.9 calculated by Kanamori~\cite{kanamori}, one obtains $-0.24(15)$~K, a value that is both small and negative, in agreement with the experimentally determined value of $-3.2(4)$~K. The close similarity between the calculated and experimentally determined values of $\theta_{\rm{CW}}$ suggests that all relevant exchange interactions have been accounted for by the $Iba2$ structure. It is important to emphasize that this analysis assumes isotropic exchange and thus assumes the isotropic part of the magnetic Hamiltonian is dominant.  While susceptibility data indicates some anisotropy, the similarity between the extracted exchange constants and the $\theta_{\rm{CW}}$ lends support for the isotropic approximation, while the slightly larger negative measured value may possibly be indicative of some anisotropic contributions. Future advances in both single crystal growth of this material and also higher flux neutron instrumentation will allow single crystal data to be obtained and the parameters refined.

\subsubsection{Single Mode Approximation \& Dimensionality} 

\begin{figure}
	\centering
	\includegraphics[width=1.0\linewidth]{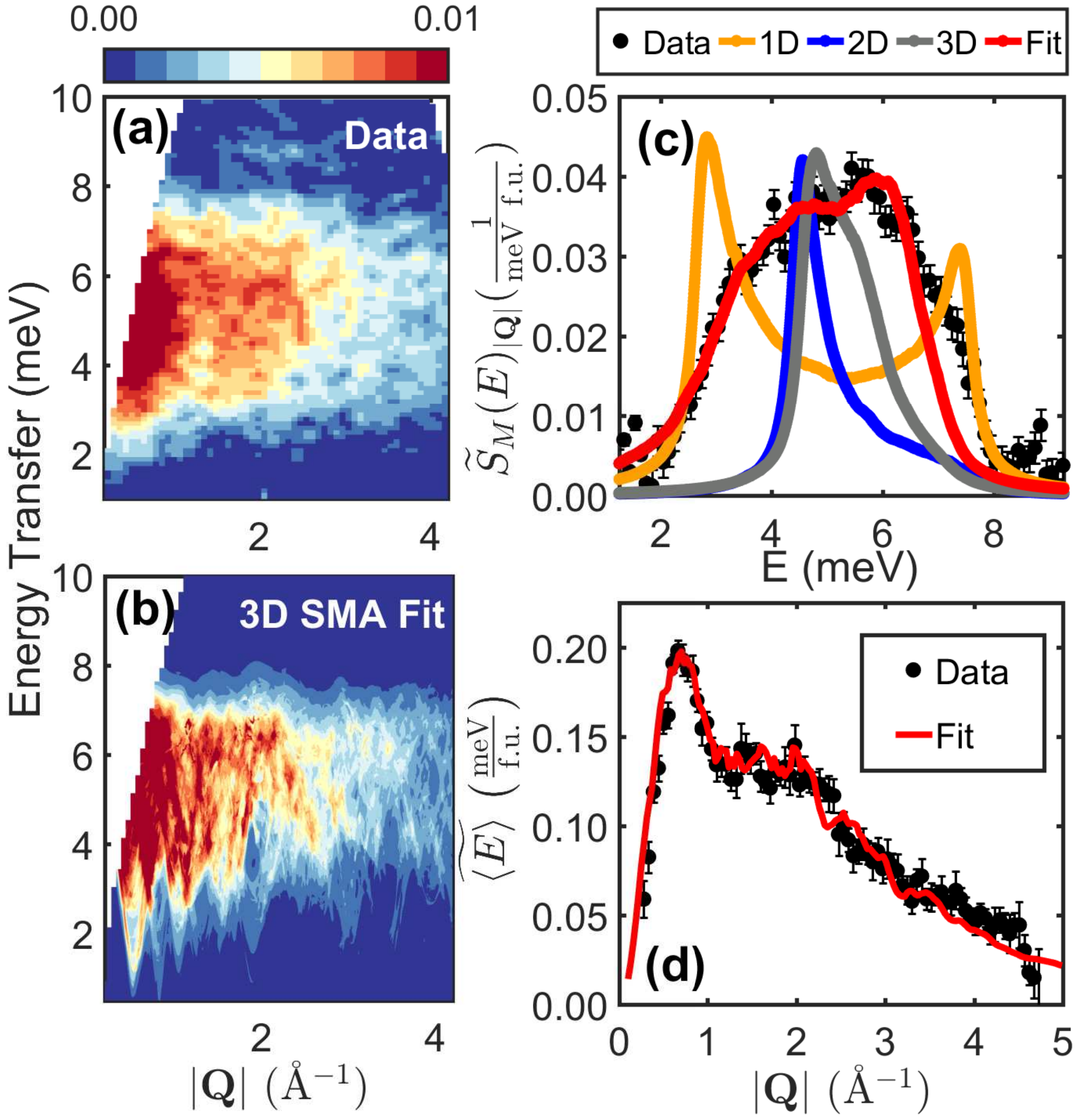}
	\caption{(a) $\widetilde{S}_{M}(|\mathbf{Q}|,E)$ measured on MARI at T=5~K with an E$\rm{_{i}}$=15~meV. (b) $\widetilde{S}_{M}(|\mathbf{Q}|,E)$ calculated by the optimization of all parameters $B_{i}$ in the heuristic model of $\epsilon(\mathbf{Q})$ in the single mode approximation of $\widetilde{S}(\mathbf{Q},E)$ utilizing the refined values of $-nJ_{ij}\langle \mathbf{S}_{i} \cdot \mathbf{S}_{j}\rangle$ from the first moment sum rule. (c) Comparison of $|\mathbf{Q}|$-integrated cuts ($|\mathbf{Q}|$=[0,3]~\AA$^{-1}$) of measured and calculated $\widetilde{S}_{M}(|\mathbf{Q}|,E)$. For the purposes of comparison, non-optimized $|\mathbf{Q}|$-integrated cuts for all three types of dimensionality $d$ are also presented. These cuts assume both $\epsilon(\mathbf{Q})$ possesses the same gap parameter $B_{o}$ obtained from the 3D SMA fit in (b) and each permissible set of parameters is equally weighted. (d) Comparison of the measured and calculated $|\mathbf{Q}|$-dependence of the first moment $\langle E \rangle$ integrated over E=[2,8]~meV.}
	\label{fig:7}
\end{figure}

\indent Since the first moment sum rule indicates the presence of multiple unique interactions spanning all three crystallographic directions in the $Iba2$ structure~\cite{znv3o8}, it was suspected that a more intriciate dispersion relation should be chosen for Eq.~\ref{eq:singlemodeapprox}, such as the expression given by 

\begin{align}
\epsilon(\mathbf{Q}) &= \left(B_{o} + B_{h}\cos(2\pi h) + B_{k}\cos(2\pi k)  + B_{l}\cos(2\pi l) \right.\nonumber\\
&\quad \left. {} + B_{hk}\{\cos[2\pi(h+k)] + \cos[2\pi(h-k)]\}  \right.\nonumber\\
&\quad \left. {} + B_{hl}\{\cos[2\pi(h+l)] + \cos[2\pi(h-l)]\}  \right.\nonumber\\
&\quad \left. {} + B_{kl}\{\cos[2\pi(k+l)] + \cos[2\pi(k-l)]\}  \right.\nonumber\\
&\quad \left. {} + B_{2h}\cos(4\pi h) + \beta_{2k}\cos(4\pi k)  + B_{2l}\cos(4\pi l)  \right)^{\frac{1}{2}},
\label{eq:dispersion}
\end{align} 

\noindent where $B_{i}$ are the dispersion parameters. The dispersion relation $\epsilon(\mathbf{Q})$ in Eq.~\ref{eq:dispersion} satisfies Bloch's theorem~\cite{squires_2012} and has been previously used to parametrize the dispersion for more complex systems involving multiple exchange interaction pathways such as PHCC~\cite{stone}, whose large dispersions could not be adequately described with the heuristic model $\epsilon(\mathbf{Q}) = \beta_{o} + \sum\limits_{i}\beta_{i}\cos(\mathbf{Q}\cdot\mathbf{d}_{ij})$~\cite{cov2o6,stone2,hong}. \\
\begin{table}
	\caption{Refined parameters of the heuristic dispersion relation in the single mode approximation of $\widetilde{S}(|\mathbf{Q}|,E)$ utilizing the refined values of $-n_{ij}J_{ij}\langle \hat{\mathbf{S}}_{i} \cdot \hat{\mathbf{S}}_{j}\rangle$ at 5~K summarized in Tab.~\ref{tab:2}. As a first approximation, the intra-plane dispersion parameters were fixed to zero. Numbers in parentheses indicate statistical errors.}
	\renewcommand{\arraystretch}{1.3} 
	\begin{tabular}{ | c | c |  } 
		\hline
		Dispersion Parameter  & Refined Value (meV$^{2}$) \\ 
		\hline
		$B_{o}$ &   28.2(3) \\ 
		\hline
		$B_{h}$ &  $-1.13(2)$ \\ 
		\hline
		$B_{k}$ &  $-4.63(4)$ \\ 
		\hline
		$B_{l}$  &  6.8(7) \\ 
		\hline
		$B_{hk}$ &  0 \\ 
		\hline
		$B_{hl}$  & 0  \\ 
		\hline
		$B_{kl}$  & 0  \\ 
		\hline
		$B_{2h}$ &  $-1.13(2)$ \\ 
		\hline
		$B_{2k}$ &  $-4.63(4)$ \\ 
		\hline
		$B_{2l}$ & 6.8(7)  \\ 
		\hline
	\end{tabular}
	\label{tab:3}
\end{table}
\indent As a first approximation, the parameters in Eq.~\ref{eq:dispersion} involving interactions between the principal axes were set to zero and each parameter along a particular principal axis was set to be equal (e.g. $B_{h} = B_{2h}$). This simple model effectively reduces Eq.~\ref{eq:dispersion} to the aforementioned simple heuristic model~\cite{cov2o6,stone2,hong} and treats every exchange interaction as a combination of interactions along the three principal axes. As illustrated in Fig.~\ref{fig:7}, all major features of $\widetilde{S}_{M}(|\mathbf{Q}|,E)$ collected at 5~K, including the large bandwidth, was successfully accounted for by a least squares optimization of the dispersion parameters. As summarized in Tab.~\ref{tab:3}, the refined dispersion parameters indicate the presence of three dimensional magnetism, consistent with the lack of significant asymmetry in the $|\mathbf{Q}|$-integrated cut $\widetilde{S}_{M}(E)_{|\mathbf{Q}|}$ displayed in Fig.~\ref{fig:7}(c), as would be expected for both 1D and 2D magnetic fluctuations~\cite{namno2,birgeneau1995,kim2001} . As summarized by Tab.~\ref{tab:3}, the dispersion parameters along $h$ and $l$ are both negative while the dispersion parameters along $k$ are positive with a larger magnitude. Both the signs and relative magnitudes of the dispersion parameters can be reconciled using the spin-flip hopping model~\cite{spinhop,cov2o6}, where $B_{i}$ for a particular direction ${i}$ is interpreted as a hopping term whose value is proportional to the energy cost of a spin-flip $t \sim SJ$ along that particular direction. The negative $h$ and $k$ dispersion parameters correspond to ferromagnetic coupling along $a$ and $b$, respectively, while the larger positive $l$ dispersion parameters correspond to stronger antiferromagnetic coupling along $c$, all consistent with both DC susceptibility and the refined magnetic structure presented in Fig.~\ref{fig:fig2}. The ability to describe the powder average magnetic dynamic response in terms of a coherent sharp mode is consistent with the cation order deduced from the critical scaling analysis and thus further evidence that the broadening of the magnetic excitations is due to powder averaging and not due to the underlying disorder.
    
\section{Discussion}
 
\subsection{Experimental limitations}     
 
\indent There are several limitations to the analysis presented in this paper. The first is the use of $\alpha$-ZnV$_{3}$O$_{8}$ as a background for the analysis of the low temperature inelastic spectrum of \cvo. As shown in Fig.~\ref{fig:1}(d), $\alpha$-ZnV$_{3}$O$_{8}$ crystallizes in the cation ordered $Iba2$ space group~\cite{znv3o8} and is thus not completely isostructural to \cvo. It can be argued that the local cation ordering deduced is an artefact of the $Iba2$ structure of the $\alpha$-ZnV$_{3}$O$_{8}$ background. To counter such a claim, we point out that the scaling analysis utilizing the same inelastic neutron scattering data, but after the subtraction of an independently calculated temperature-independent background derived from detailed balance~\cite{background,hong}, provided a critical exponent $\nu$ consistent with pure 3D Ising behavior. Such pure 3D Ising behavior would be unexpected if Co$^{2+}$ was locally disordered. \\
 \indent Another limitation is the observation that the low temperature cooperative magnetism of \cvo~can be treated as exclusively due to coupling between Co$^{2+}$ moments. The presence of a second magnetic disordered ``counter''-cation is in contrast to the model dilute 3D Ising antiferromagnets where the ``counter"-cations are non-magnetic and thus interactions between magnetic ions of one type (e.g. Fe$^{2+}$) are exclusively considered~\cite{young1998,raposo,cowley1972properties,Cowley1984,birgeneau87:61}. Such a situation was assumed to apply to \cvo~in the analysis presented so far as a first approximation since there is evidence that V$^{4+}$ behaves paramagnetically; but it is highly unlikely that coupling between V$^{4+}$ and other V$^{4+}$ or Co$^{2+}$ plays no role in the low temperature magnetism and thus the analogy to the dilute antiferromagnets such as Fe$_{x}$Zn$_{1-x}$F$_{2}$ should be approached with caution. It is important to note that the apparent lack of influence of V$^{4+}$ coupling, relative to coupling between Co$^{2+}$ cations, may be due to the exclusive use of $t_{2g}$ orbitals by V$^{4+}$, in contrast to the $e_{g}$ orbitals utilized by Co$^{2+}$ which is predicted to give much stronger coupling~\cite{rules1,rules2,rules3,projection,fiete}. \\
 \indent A further limitation concerns the nature of the competing ferromagnetic and antiferromagnetic interactions in \cvo. In contrast to the Fe$_{x}$Zn$_{1-x}$F$_{2}$ series~\cite{inelastic,king1981,cowley1972properties}, \cvo~exhibits both distinct ferromagnetic inter-chain and antiferromagnetic intra-chain coupling along the $ab$ plane and along $c$, respectively. Both ferromagnetic and antiferromagnetic coupling possess similar magnitudes as proven by their near cancellation corresponding to a Weiss temperature near zero.  With a Weiss temperature near zero, combined with a $T\rm{_{N}}\sim$~19~K, the frustration index $f = \left|\frac{\theta_{CW}}{T_{N}} \right|\lesssim~1$ implies the absence of frustration, a key contributor to the rich phase diagram of the dilute 3D Ising antiferromagnets~\cite{raposo}. To address the concurrent presence of both ferromagnetic and antiferromagnetic couplings, it is worth noting that such a situation is reminiscent of another random dilute 3D Ising magnet system Fe$_{x}$Mg$_{1-x}$Cl$_{2}$ where $x >$ 0.55, a series of compounds whose magnetic properties have been shown consistently to be qualitatively similar to that of Fe$_{x}$Zn$_{1-x}$F$_{2}$~\cite{matteson1994,Fytas2018}. To address the absence of frustration, it is worth noting that in contrast to the current study, previous work~\cite{oka} on smaller hydrothermally grown crystals of \cvo~reported a $T\rm{_{N}}=$~8.2~K and a Weiss temperature of $-32.1$~K, corresponding to a frustration index $f\sim$~4, indicating evidence for significant frustration. Such contrasting behavior provides strong evidence that sample dependence may play a significant role in determining the magnetic properties of \cvo, as has been consistently observed for the dilute antiferromagnets, whose response functions are significantly influenced by both sample quality and non-equilibrium physics~\cite{fisher,king1981,cowley1972properties}. The particular dependence on sample quality can be partially rationalized using recent work by Volkova~\cite{volkova} on $\alpha$-ZnV$_{3}$O$_{8}$. Numerical simulations indicated that although the ordered-$Iba2$ arrangement was predicted to exhibit minimal frustration, if one instead assumed a disordered-$Ibam$ arrangement, significant magnetic frustration was predicted to manifest itself as competing inter-chain couplings of similar magnitudes in the presence of a dominant antiferromagnetic intra-chain coupling. The contrasting behavior between $Iba2$ and $Ibam$ cationic arrangement may provide an explanation for the aforementioned difference in the experimentally determined frustration indices with samples possessing more disorder exhibiting a larger value of $f$. 
 
 \subsection{Disordered \emph{Ibam} versus ordered \emph{Iba2}?}
 
\indent A contradiction arises from a combined analysis of x-ray and neutron diffraction, DC susceptibility and inelastic neutron spectroscopy measurements in that the disordered-$Ibam$ structure is derived from diffraction measurements, however the dynamics are more consistent with an ordered-$Iba2$ arrangement of Co$^{2+}$ ions.  Diffraction indicates that statically the arrangement of Co$^{2+}$ ions is disordered, however the collective long wavelength fluctuations seem to average out this disorder. \cvo~therefore appears to be magnetically ordered for longer lengthscales.  The disorder in $\alpha$-CoV$_{3}$O$_{8}$ differs from a Griffiths phase where local order is present and maybe more analogous to the situation in water ice where local selection rules are present.~\cite{Keen15:521}  However, the lack of strong diffuse scattering in our single crystal experiments makes a comparison to these correlated disordered systems difficult.   However, the presence of the local selection structural selection for Co$^{2+}$ and V$^{4+}$ distinguishes CoV$_{3}$O$_{8}$ from a doped random magnet where no such local order is required.  The apparent robustness of \cvo~to disorder is discussed below in the context of spin-orbit coupling and comparison to other model magnets in a random field.  

 \subsection{Universality class of \cvo}
 
\indent Ising anisotropy is experimentally supported by several observations discussed above: the presence of a significant octahedral distortion ($\delta\sim$~11) as deduced from a combination of single crystal x-ray and neutron diffraction data, the presence of 3D Ising fluctuations as deduced from both critical exponents $\nu$ and $\beta$, and the presence of strong spin-orbit coupling supported by neutron spectroscopy.  3D dimensionality ($d$~=~3) is suggested based on the following: the values of the critical exponents $\nu$ and $\beta$, the non-zero refined values of all $h$, $k$ and $l$ dispersion parameters in $\epsilon(\mathbf{Q})$ reflecting both strong coupling  in both the $ab$ plane and along $c$, in combination with the relatively weak anisotropy of the DC susceptibility. \\
\indent The random magnetic cation distribution is supported by the refined $Ibam$ structure from both single crystal x-ray and neutron diffraction and the value of $\beta$. An additional observation is the intrinsic width of the AFM transition as measured with DC susceptibility, reflected by the large experimental error of $\beta$ caused by the rounding of the order parameter measurement, as has been experimentally observed in other dilute 3D Ising antiferromagnets such as Co$_{x}$Zn$_{1-x}$F$_{2}$~\cite{cowley_Co}.  The dilution of 3D Ising magnetism can be rationalized by the key observation that V$^{4+}$ appears to remain purely paramagnetic down to 2~K and thus has no significant influence on the low temperature cooperative magnetic properties of \cvo, as proven by a combination of inelastic neutron scattering and DC susceptibility measurements. 
 
 \subsection{Comparison between \cvo~and Random Field Ising magnets}    
 
If one disregards the magnetic influence of V$^{4+}$, effectively treating the cation as a ``counter"-ion such as Zn$^{2+}$ in Fe$_{x}$Zn$_{1-x}$F$_{2}$ or Mn$_{x}$Zn$_{1-x}$F$_{2}$, then the magnetism due to Co$^{2+}$ in \cvo~may be regarded as being magnetically diluted by 50\%. Additionally, it is important to note that the failure to observe strong structural diffuse scattering with x-ray and neutron diffraction measurements is suggestive of a lack of local cation ordering or gradients.  These concentration gradients were noted in dilute model antiferromagnets~\cite{fisher,king1981,young1998,ariano1982,ramos1988,belanger1988}. Such a combination of significant dilution and disorder would be expected to have a significant effect on the dynamics~\cite{belanger1993,uemura,inelastic,terao1994,fisher,vojta2013,vojtabook,vojta13:1550,vojta06:39,hoyos08:100}.  In this sense, it is surprising that there seems to be little effect on the magnetic dynamics in \cvo, where the magnetic excitations are consistent with a fully ordered cation arrangement. Such behavior is suggestive that hydrodynamic and long wavelength fluctuations are not strongly sensitive to the disorder in \cvo, in contrast with expectations based on theory~\cite{korenbilt,wan93:48}. The robust nature of the dynamics to dilution, and in particular disorder is analogous to several observations in dilute random field magnets and in particular the Fe$_{x}$Zn$_{1-x}$F$_{2}$ series~\cite{rodriguez2007,hutchings1970,araujo1980}, where sharp excitations are still observable for large amount of doping~\cite{inelastic}.  Unlike members of the Fe$_{x}$Zn$_{1-x}$F$_{2}$ series closer to the percolation threshold ($x_{p}\sim$~0.24) that exhibit spin glass behavior~\cite{inelastic,raposo,belanger1993,alvarez2012}, Fe$_{0.5}$Zn$_{0.5}$F$_{2}$ assumes long range antiferromagnetic order in zero field with a $T\rm{_{N}}$ corresponding to half of that of FeF$_{2}$~\cite{birgeneau1983a,hill1997}.  The appearance of long range antiferromagnetic order as measured by DC susceptibility with a $\mu_{o}H_{ext}$~=~0.5~T supports the claim that \cvo~is not close to the percolation threshold, where even the smallest external field destroys long range order, as is the case for  M$_{x}$Zn$_{1-x}$F$_{2}$, where M~=~Co$^{2+}$ and Fe$^{2+}$ ~\cite{hagen}.   However, the random field Ising magnet Mn$_{x}$Zn$_{1-x}$F$_{2}$~\cite{mitchell,cluster} does show strong effects of the disorder on the dynamics. Such behavior is consistent with cases of random fields introduced through confinement where when the critical fluctuations have a similar length scale to the underlying disorder, the phase transition is strongly altered~\cite{fisher,Chan96:49,hoyos08:100,vojtabook,vojta2013}. \\
\indent A key difference between MnF$_{2}$ and both \cvo~and FeF$_{2}$ is the presence of strong crystal field effects and spin-orbit coupling in the latter two compounds~\cite{Hutchings72:5,hutchings1970}. It is also worth noting that unlike the case of pure CoO~\cite{kanamori,yamani08:403,cowley,sarte18:98,satoh17:8,tomiyasu2006,feygenson11:83} where the large and far reaching exchange constants result in a significant and ultimately problematic entanglement of spin-orbit levels~\cite{sarte18:98}, in the case of \cvo, the exchange constants are weak and the Weiss temperature is near 0~K. Both observations suggest that the presence of both strong crystal field effects and spin-orbit coupling with well-separated $j\rm{_{eff}}$ manifolds, as is the case for \cvo, may be central to making the dynamics robust against strong disorder.

\section{Concluding Remarks} 

\indent In summary, a combination of zero field diffraction, DC susceptibility and neutron spectroscopy measurements have indicated that the low temperature cooperative magnetism of \cvo~is dominated by $j\rm{_{eff}}=\frac{1}{2}$ Co$^{2+}$ cations randomly distributed over the 16$k$ metal site of the $Ibam$ structure, thus corresponding to an intrinsically disordered magnet without the need for any external influences such as chemical dopants or porous media. Despite the intrinsic disorder, by employing the sum rules of neutron scattering, the collective excitations have been shown to not be significantly affected by the disorder, displaying behavior consistent with an ordered-$Iba2$ arrangement of $j\rm{_{eff}}=\frac{1}{2}$ Co$^{2+}$ moments over a macroscopic scale. These Co$^{2+}$ moments are coupled $via$ a 3D network of competing  ferromagnetic and stronger antiferromagnetic superexchange interactions within the $ab$ plane and along $c$, respectively, resulting in long range antiferromagnetic order of the Co$^{2+}$ moments at $T\rm{_{N}}\sim$19~K, despite a Weiss temperature near 0 K.  A comparison of our results to the random 3D Ising magnets and other compounds where spin-orbit coupling is present indicate that both the presence of an orbital degree of freedom, in combination with strong crystal field effects and well-separated $j\rm{_{eff}}$ manifolds may be key in making the dynamics robust against disorder.

\section{Acknowledgements} 

\indent We acknowledge discussions with J.A.M.~Paddison, C.R.~Wiebe, A.J.~Browne, G.~Perversi, S.E.~Maytham (HBS) and D.R.~Jarvis. We are grateful to the Royal Society, the STFC, the ERC and the EPSRC for financial support. P.M.S. acknowledges financial support from the CCSF, the RSC and the University of Edinburgh through the GRS and PCDS. We acknowledge the support of the National Institute of Standards and Technology, US Department of Commerce, in providing a portion of the neutron research facilities used in this work.  Finally, the authors would like to thank the Carnegie Trust for the Universities of Scotland for providing facilities and equipment for chemical synthesis. 

\newpage

\onecolumngrid
\appendix
\section{Crystallographic data}\label{sec:appendixA} 
\begin{table*}[h]
	\caption{Crystal data, experimental and structural refinement parameters for single crystal x-ray diffraction measurements on \cvo. Numbers in parentheses indicate statistical errors.}
	\begin{tabular}{ | c | c |} 
		\hline
		Parameter &  Value  \\ 
		\hline
		Empirical Formula & CoV$_{3}$O$_{8}$ \\
		\hline 
		Formula weight & 339.7529 g~mol$^{-1}$ \\
		\hline
		Temperature & 120.0(1)~K \\
		\hline
		Crystal Dimensions & 0.40~$\times$~0.11~$\times$~0.09~mm$^{3}$ \\
		\hline
		Wavelength & 0.71073~\AA (Mo K$_{\alpha}$) \\
		\hline
		Crystal System & Orthorhombic \\
		\hline
		Space Group & $Ibam$ (\#72) \\
		\hline
		$a$	 & 14.29344(4) \AA  \\ 
		\hline
		$b$ & 9.8740(3) \AA \\
		\hline
		$c$ & 8.34000(3) \AA \\
		\hline 
		$V$ & 1185.60(6) \AA$^{3}$ \\
		\hline 
		$Z$ & 8 \\ 
		\hline 
		$\rho$ & 3.8069(3) g~cm$^{-3}$\\
		\hline
		$\theta$ range for data collection & 4.13$\rm{^{\circ}}~\leq~\theta~\leq$ 30.18$\rm{^{\circ}}$ \\ 
		\hline
		Limiting Indices & $-19~\leq~h~\leq~20$, $-13~\leq~k~\leq~14$ and $-11~\leq~l~\leq~11$ \\
		\hline 
		Number of Reflections $I > 0$ & 985 \\
		\hline 
		Number of Reflections $I > 3\sigma(I)$ & 910 \\
		\hline
		Absorption Correction Method & Gaussian \\ 
		\hline
		Extinction Method & B-C Type 1 Gaussian Isotropic \\ 
		\hline
		Extinction Coefficient & 2300(100)\\
		\hline  
		Refinement Method & Full matrix least squares on $F^{2}$\\
		\hline
		Number of Parameters(Constraints)  & 67(9) \\
		\hline
		R$\rm{_{F^{2}}}$ ($I > 3\sigma(I)$, all) & 1.65\%, 1.90\% \\
		\hline
		R$\rm{_{wF^{2}}}$ ($I > 3\sigma(I)$, all) & 2.38\%, 2.46\%  \\
		\hline
		Goodness-of-Fit $\chi^{2}$ ($I > 3\sigma(I)$, all) & 1.47\%, 1.48\% \\
		\hline
	\end{tabular}
	\label{tab:ap1}
\end{table*}

\begin{table*}[htb]
	\caption{Structural parameters of \cvo~obtained from  the refinement of single crystal x-ray diffraction data collected at 120~K. Numbers in parentheses indicate statistical errors.}
	\begin{tabular}{ | c | c |c|c|c|c|c|} 
		\hline
		Atom (Label) & Wyckoff Position & $x$ & $y$ & $z$ & $B_{iso}$ (\AA$^{2}$) & Fractional Occupancy \\
		\hline
		Co & 16$k$  & 0.654760(16)  & 0.33285(2)  & 0.81060(3)  & 0.39(2)  &  0.506(6)\\
		\hline
		V(1) & 16$k$  & 0.654760(16)  & 0.33285(2)  & 0.81060(3)  & 0.39(2) & 0.494(6) \\
		\hline
		V(2) & 8$j$ & 0.52271(2)  & 0.16672(4)  & 0.5  & 0.321(5)  & 1  \\
		\hline
		V(3) & 8$j$  & 0.70168(2)  & 0.94348(4)  & 0.5  & 0.252(6) & 1 \\
		\hline
		O(1) & 8$j$  & 0.73349(11)  & 0.41325(16)  & 0  & 0.52(2) &  1\\
		\hline
		O(2) & 8$j$  & 0.58248(10)  & 0.27500(16)  & 0 & 0.50(2)  & 1\\
		\hline
		O(3) & 16$k$ & 0.76787(8) & 0.35258(11)  & 0.66386(15)  & 0.53(2) &  1\\
		\hline
		O(4) & 8$f$  & 0.61080(11)  & 0.5  & 0.75  & 1.2(1) &  1\footnote{The assignment of full occupancy in the 8$f$ position corresponding to the bridging oxygen is in agreement with the initial refinement by Oka \emph{et al.}~\cite{oka}.}\\
		\hline
		O(5) & 16$k$ & 0.57900(8) & 0.22361(12)  &  0.65802(16) & 0.79(3) &  1\\
		\hline
		O(6) & 8$j$  & 0.57973(10) & 0.98272(16)  & 0.5  & 0.48(2) &  1\\
		\hline
	\end{tabular}
	\label{tab:ap2}
\end{table*}

\begin{table*}[htb]
	\caption{Crystal data, experimental and structural refinement parameters for single crystal neutron diffraction measurements on \cvo. Numbers in parentheses indicate statistical errors.}
	\begin{tabular}{ | c | c |} 
		\hline 
		Parameter &  Value  \\
		\hline
		Empirical Formula & CoV$_{3}$O$_{8}$ \\
		\hline 
		Formula weight & 339.7529 g~mol$^{-1}$ \\
		\hline
		Temperature & 5.00(3)~K \\
		\hline
		Crystal Dimensions & 13.2~$\times$~4.1~$\times$~2.1~mm$^{3}$ \\
		\hline
		Wavelength & Polychromatic (time-of-flight) \\
		\hline
		Crystal System & Orthorhombic \\
		\hline 
		Nuclear Space Group & $Ibam$ (\#72) \\
		\hline
		Magnetic Space Group (BNS Setting) & $P_{I}ccn$ (\#56.376) or ($I_{P}bam'$ OG 72.10.639) \\
		\hline
		$\mathbf{k}$ & (111) \\  
		\hline
		$a$	 & 14.3280(4) \AA \\ 
		\hline
		$b$ & 9.9213(3) \AA  \\ 
		\hline
		$c$ &  8.4160(3) \AA \\ 	
		\hline
		$V$ & 1196.35(7) \AA$^{3}$ \\ 
		\hline 
		$Z$ & 8 \\
		\hline 
		$\rho$ & 3.773(3) g~cm$^{-3}$\\ 
		\hline 
		$\theta$ range for data collection & 2.94$\rm{^{\circ}}~\leq~\theta~\leq$ 76.22$\rm{^{\circ}}$ \\ 
		\hline
		Limiting Indices & $-35~\leq~h~\leq~33$, $-25~\leq~k~\leq~19$ and $-16~\leq~l~\leq~22$ \\
		\hline 
		Number of Reflections $I > 0$ & 5120  \\ 
		\hline 
		Number of Reflections $I > 3\sigma(I)$ &  5086 \\ 
		\hline
		Refinement Method & Full matrix least squares on $F^{2}$\\
		\hline
		Absorption Correction & None \\ 
		\hline
		Extinction Method & B-C Type 1 Gaussian Isotropic \\ 
		\hline
		Extinction Coefficient & 348(8)\\
		\hline  
		Number of Parameters(Constraints)  &  34(10) \\ 
		\hline
		$\mu_{a}$& 1.35(4) $\rm{\mu_{B}}$   \\ 
		\hline
		$\mu_{b}$	& $1.16(5)$ $\rm{\mu_{B}}$  \\ 
		\hline
		$\mu_{c}$	& 3.05(4) $\rm{\mu_{B}}$  \\  
		\hline
		R$\rm{_{F^{2}}}$ ($I > 3\sigma(I)$, all) & 8.34\%, 8.38\%  \\ 
		\hline
		R$\rm{_{wF^{2}}}$ ($I > 3\sigma(I)$, all) & 8.98\%, 8.99\%  \\ 
		\hline
		R$\rm{_{F^{2}_{mag}}}$ ($I > 3\sigma(I)$, all) & 23.44\%, 24.13\% \\ 
		\hline
		Goodness-of-Fit $\chi^{2}$ ($I > 3\sigma(I)$, all)	& 3.18, 3.19 \\ 
		\hline
	\end{tabular}
	\label{tab:ap3}
\end{table*}

\begin{table*}[htb]
	\caption{Structural parameters for the nuclear structure of \cvo~obtained from the refinement of single crystal neutron diffraction data collected at 5~K. Numbers in parentheses indicate statistical errors.}
	\begin{tabular}{ | c | c |c|c|c|c|c|} 
		\hline
		Atom (Label) & Wyckoff Position & $x$ & $y$ & $z$ & $U_{iso}$ (\AA$^{2}$) & Fractional Occupancy\footnote{The value of the fractional occupancies were fixed to the refined values obtained from a refinement of single crystal neutron diffraction data collected at 50~K.} \\
		\hline
		Co & 16$k$  & 0.9068(3)  & 0.5765(3)  & 1.0616(5)  & 0.0005(5)  & 0.504(4)  \\
		\hline
		V(1) & 16$k$  & 0.9068(3)  & 0.5765(3)  & 1.0616(5) & 0.0005(5)   & 0.496(4) \\
		\hline
		V(2) & 8$j$ & 0.771  & 0.416  & 0.75  & 0.0042  & 1 \\
		\hline
		V(3) & 8$j$  & 0.957  & 1.198  & 0.75 & 0.0042   & 1 \\
		\hline
		O(1) & 8$j$ & 0.98357(9)  & 0.66282(11)   & 0.25  & 0.00356(16)   & 1 \\
		\hline
		O(2) & 8$j$ & 0.83242(8)  & 0.52523(11)   & 0.25  & 0.00388(16)   & 1 \\
		\hline
		O(3) & 16$k$ & 1.01787(6)  & 0.60246(8)  & 0.91384(10)  & 0.00392(11)  & 1 \\
		\hline
		O(4) & 8$f$  & 0.86076(9)   & 0.75   & 1  & 0.0126(4)  & 1 \\
		\hline
		O(5) & 16$k$ & 0.82899(6)  & 0.47373(8)  & 0.90801(10)  & 0.00556(12)  &  1 \\
		\hline
		O(6) & 8$j$  & 0.82973(8)   & 1.23252(10)  & 0.75  & 0.00269(15)   & 1 \\
		\hline
	\end{tabular}
	\label{tab:ap4}
\end{table*}

\begin{table*}[hbt]
	\caption{Cell parameters, fit residuals and agreement factors for \cvo~obtained from the Rietveld refinement of laboratory powder x-ray diffraction data collected at 300~K. Numbers in parentheses indicate statistical errors.}
	\begin{tabular}{ | c | c |} 
		\hline
		Parameter & Value   \\ 
		\hline
		$a$	 & 14.292(1)~\AA     \\ 
		\hline
		$b$	 & 9.8844(9)~\AA     \\ 
		\hline
		$c$	 & 8.3969(8)~\AA      \\ 
		\hline
		$V$	 &1186.2(3)~\AA$^{3}$      \\ 
		\hline
		$\chi^{2}$	 & 1.487    \\ 
		\hline
		$\rm{R_{p}}$	 & 10.26\%    \\ 
		\hline
		$\rm{R_{wp}}$	 & 14.05\%    \\ 
		\hline		
	\end{tabular}
\end{table*} 

\begin{table*}[htb]
	\caption{Cobalt-oxygen distances and corresponding octahedral distortion parameter~\cite{cov2o6,delta} $\delta$ for \cvo~at 5~K deduced from the Rietveld refinement of single crystal neutron diffraction data. Numbers in parentheses indicate statistical errors.}
	\begin{tabular}{ | c | c | c| } 
		\hline
		Oxygen Label & $d$ (\AA) & $\left(\frac{d - \langle d \rangle}{\langle d \rangle} \right)^{2} \times 10^{4}$  \\ 
		\hline
		O(1)	 & 2.12527(5) & 27.79(3)   \\ 
		\hline
		O(2)	 & 1.98916(5) & 2.162(8) \\ 
		\hline
		O(3)	 & 2.07872(5) & 8.79(2) \\
		\hline
		O(3)$'$     & 2.02358(5)  & 0.0549(13)   \\ 
		\hline
		O(4)	 & 1.92213(5)  & 22.95(3)\\ 
		\hline
		O(5)	 & 1.97422(4)  & 4.89(1) \\
		\hline
		\multicolumn{2}{|c|}{$\frac{1}{N}\sum \left\{ \left(\frac{d - \langle d \rangle}{\langle d \rangle}\right)^{2} \times 10^{4} \right\}$} & 11.106(8) \\ 
		\hline
	\end{tabular}
\end{table*}

\begin{figure*}[htb]
	\centering
	\includegraphics[width=1\linewidth]{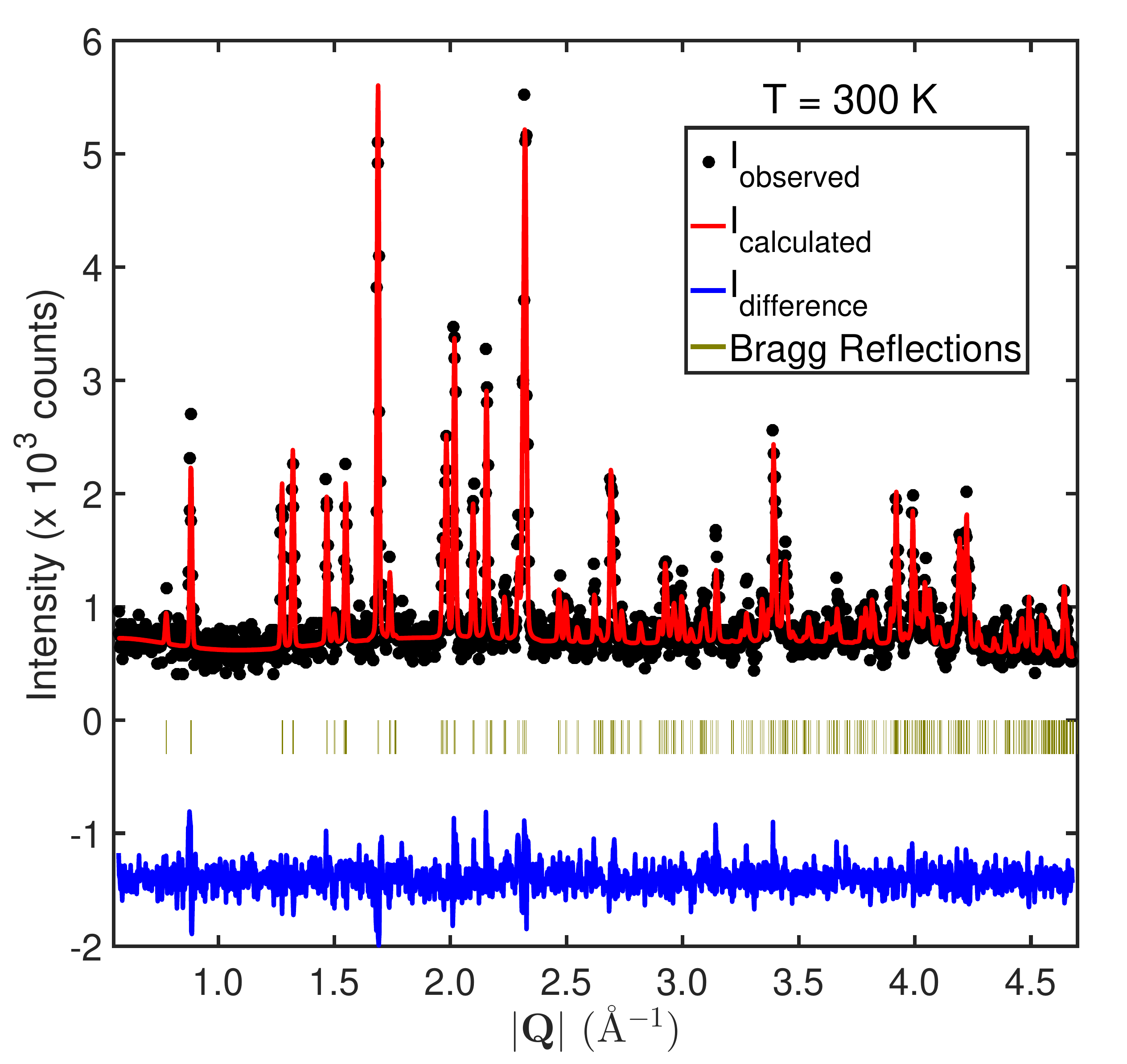}
	\caption{Room temperature diffraction profile of polycrystalline \cvo~collected on a Bruker D2 Phaser x-ray diffractometer utilizing a monochromated Cu K$_{\alpha,1,2}$ source, confirming the absence of any discernible impurities. A Rietveld refinement ($\chi^{2}$~=~1.487, $R_{p}$~=~10.26\%, $R_{wp}$~=~14.05\%) indicates \cvo~crystallizes in the orthorhombic $Ibam$ (S.G. \#72) structure ($a$~=~14.292(1)~\AA, $b$~=~9.8844(9)~\AA, $c$~=~8.3969(8)~\AA).}
	\label{fig:powder}
\end{figure*}
\clearpage
\twocolumngrid
\section{Projection Factors}  

\indent As outlined in the main text, a comparison between the current study and previous studies on other Co$^{2+}$-based magnets~\cite{ross,sarte18:98,buyers,cowley,kanamori,cowley_Co,cov2o6} suggests that the low temperature magnetism of \cvo~may be solely attributed to the ground state doublet spin-orbit manifold, and thus can be simplified to a $j=\frac{1}{2}$ model. To utilize such a model in the current study, the Land\'{e} $g$-factor $g_{J}$ was required to be projected onto individual $j \equiv j\rm{_{eff}}$ manifolds.  

\subsection{Calculation of the Orbital Angular Momentum Operator Projection Factor $\alpha$} \label{sec:projection} 

\indent Before proceeding with the projection of the Land\'{e} $g$-factor onto the $j\rm{_{eff}}=\frac{1}{2}$ ground state spin-orbit manifold, it is important to note that such a doublet manifold is a consequence of an approach commonly used~\cite{cowley,buyers,cov2o6,sarte18:98,kanamori} to address the orbital triplet ground state in Co$^{2+}$. Such an approach first defines an effective total angular momentum $\hat{\mathbf{j}}\rm{_{eff}} = \hat{\mathbf{l}} + \hat{\mathbf{S}}$, where $\hat{\mathbf{l}}$ is a fictitious orbital angular momentum operator, with eigenvalue $l=$1 to reflect an triplet orbital degeneracy~\cite{abragam}. Thus, a projection of $g_{J}$ and any angular momentum operators onto the $j=\frac{1}{2}$ manifold requires a concurrent projection of $\hat{\mathbf{L}}$ onto  $\hat{\mathbf{l}}$, \emph{via} a projection factor $\alpha$.  \\ 
\indent The determination of the projection factor $\alpha$ begins by first defining the crystal field Hamiltonian $\hat{\mathcal{H}}_{CEF}$ describing the effects of the crystalline electric field on the free ion states of the $d^{7}$ Co$^{2+}$ resulting from the symmetry imposed by the crystal lattice~\cite{cowley,kanamori,buyers,seitz1959}. Assuming both negligible distortions away from purely octahedral coordination and negligible admixture between the $^{4}F$ ground and first excited $^{4}P$ free ion states, a weak crystal field approach can be employed~\cite{hutchings} whereby $\hat{\mathcal{H}}_{CEF}$ can be written in terms of the Stevens operators $\hat{\mathcal{O}}^{0}_{4}$, $\hat{\mathcal{O}}^{4}_{4}$ and the numerical coefficient $B_{4}$ as 

\begin{equation}
\hat{\mathcal{H}}_{CEF} = B_{4}\left(\hat{\mathcal{O}}^{0}_{4} + 5\hat{\mathcal{O}}^{4}_{4}\right). 
\label{eq:alpha1}
\end{equation}     

\noindent The numerical coefficient $B_{4}$ is defined as $\beta\langle r^{4} \rangle$ where $\beta$ is the Stevens multiplicative factor, while the Stevens operators are defined in terms of the $\hat{L}^{2}$, $\hat{L}_{z}$ and $\hat{L}_{\pm}$ orbital angular momentum operators~\cite{hutchings,abragam} as 

\begin{equation}
\hat{\mathcal{O}}^{0}_{4}=35\hat{L}_{z}^{4}-30\hat{L}^{2}\hat{L}_{z}^{2}+25\hat{L}_{z}^{2}-6\hat{L}^{2}+3\hat{L}^{4},
\label{eq:alpha2}
\end{equation} 

\noindent and 

\begin{equation}
\hat{\mathcal{O}}^{4}_{4}=\frac{1}{2}\left[\hat{L}_{+}^{4} + \hat{L}_{-}^{4} \right].
\label{eq:alpha3}
\end{equation} 

\indent By combining Eqs.~\ref{eq:alpha1}-\ref{eq:alpha3} and setting $B_{4}$ as $-1$, the crystal field Hamiltonian is given by 

\begin{equation}
\left[ {\begin{array}{ccccccc}
	 -180 & 0  & 0 & 0 & -232.4 & 0 & 0 \\
	0 & 420 & 0 & 0 & 0 & -300 & 0 \\
	 	0 & 0 & -60 & 0 & 0 & 0 & -232.4  \\
	 0 & 0 & 0 & -360 & 0 & 0 & 0  \\
	 -232.4 & 0 & 0 & 0 & -60 & 0 & 0\\
	 0 &  -300 & 0 & 0 & 0 & 40 & 0  \\
	 	 0 & 0 & -232.4 & 0  & 0 & 0 & -180 \\
	\end{array} } \right]
\label{eq:alpha4}
\end{equation}  

\noindent in the $|L=3,m_{L}\rangle$ basis where each operator has been normalized by $\hbar$. Before proceeding, it is worth noting that by setting $|B_{4}|$ as 1, all energy eigenstates will be in terms of $B_{4}$ while the negative sign is due to the $d^{7}$ electron configuration of Co$^{2+}$, producing a triplet and not a singlet ground state like Ni$^{2+}$~\cite{seitz1959,kim11:84}.\\
\indent Diagonalizing the crystal field Hamiltonian yields 

\begin{equation} 
\left[ {\begin{array}{ccccccc}
	\textcolor{model7}{-360} & 0  & 0 & 0 & 0& 0 & 0 \\
		0 & \textcolor{model7}{-360}  & 0 & 0 & 0& 0 & 0 \\
			0 & 0  & \textcolor{model7}{-360} & 0 & 0& 0 & 0 \\
				0 & 0  & 0 & \textcolor{model6}{120} & 0& 0 & 0 \\
					0 & 0  & 0 & 0 & \textcolor{model6}{120}& 0 & 0 \\
						0 & 0  & 0 & 0 & 0& \textcolor{model6}{120} & 0 \\
							0 & 0  & 0 & 0 & 0& 0 & \textcolor{model2}{720} \\
	\end{array} } \right]
\label{eq:alpha5}
\end{equation} 

\noindent corresponding to a triply degenerate \textcolor{model7}{ground state} ($\Gamma_{4}$), a triply degenerate \textcolor{model6}{first excited state} ($\Gamma_{5}$) and a \textcolor{model2}{singlet second excited state} ($\Gamma_{2}$), where $\Delta(\Gamma_{4} \rightarrow \Gamma_{5})=480B_{4}$ and  $\Delta(\Gamma_{5} \rightarrow \Gamma_{2})=600B_{4}$. \\
\indent Utilizing the diagonalized crystal field Hamiltonian above, a transformation matrix $\mathcal{C}$ can be defined as 

\begin{equation}
\mathcal{C} = \left[ {\begin{array}{ccccccc}
	0 & 0  & -0.79 & 0.61 & 0& 0 & 0 \\
	0 & 0  & 0 & 0 & -0.71 & 0 & -0.71 \\
	0.61 & 0  & 0 & 0 & 0 & -0.79  & 0 \\
	0 & 1.00  & 0 & 0 & 0& 0 & 0 \\
	0 & 0  & -0.61 & -0.79 & 0& 0 & 0 \\
	0 & 0  & 0 & 0 & -0.71 & 0 & 0.71 \\
	0.79 & 0  & 0 & 0 & 0 & 0.61 & 0 \\
	\end{array} } \right]
\label{eq:alpha6}
\end{equation}

\noindent where the columns of $\mathcal{C}$ are the eigenvectors corresponding to the eigenvalues in Eq.~\ref{eq:alpha5}. The eigenvectors are arranged in the order of increasing eigenvalues from left to right. In the case of degenerate eigenvalues, the eigenvectors are arranged in the order of increasing eigenvalues from left to right after the application of a small perturbative magnetic field $\hat{\mathcal{H}}_{MF} = H_{MF}\hat{S}_{z}$. The transformation matrix $\mathcal{C}$ rotates operators from the $|L=3,m_{L}\rangle$ basis to a  $|\phi\rm{_{CEF}}\rangle$ basis defined by the crystal field eigenvectors by

\begin{equation}
\hat{\mathcal{O}}_{|\phi_{CEF}\rangle} = \mathcal{C}^{-1}\hat{\mathcal{O}}_{|L,m_{L}\rangle}\mathcal{C}.
\label{eq:transform}
\end{equation} 

\indent Since the ground state multiplet of the crystal field Hamiltonian corresponds to the triply orbitally degenerate manifold, then the \textcolor{model7}{top} $3 \times 3$ block matrix of the $z$-component of the orbital angular momentum operator projected onto the crystal field basis must~\cite{fiete}: (1) have its matrix entries arranged in a format equivalent to its corresponding angular momentum operator with $l=$~1, while (2) the entries in both matrices must be equal up to the projection constant $\alpha$. Projecting the $\hat{L}_{z}$ operator from the $|L=3,m_{L}\rangle$ basis to the $|\phi\rm{_{CEF}}\rangle$ basis \emph{via} Eq.~\ref{eq:transform}, one obtains

\begin{equation}
\mathcal{C}^{-1}\hat{L}_{z}\mathcal{C} = \left[ {\begin{array}{ccc|ccc|c}
	\textcolor{model7}{1.50} & \textcolor{model7}{0}  & \textcolor{model7}{0} & 0 & 0& -1.94 & 0 \\
\textcolor{model7}{0} & \textcolor{model7}{0}  & \textcolor{model7}{0} & 0 & 0& 0 & 0 \\
\textcolor{model7}{0} & \textcolor{model7}{0}  & \textcolor{model7}{-1.50} & -1.94 & 0& 0 & 0 \\
\hline
0 & 0  & -1.94 & 	\textcolor{model6}{-0.50} & 	\textcolor{model6}{0}& 	\textcolor{model6}{0} & 0 \\
0 & 0  & 0 & 	\textcolor{model6}{0} & 	\textcolor{model6}{0} & 	\textcolor{model6}{0} & 2.00 \\
-1.94 & 0  & 0 & 	\textcolor{model6}{0} & \textcolor{model6}{0}& \textcolor{model6}{0.50} & 0 \\
\hline
0 & 0  & 0 & 0 & 2.00& 0 & \textcolor{model2}{0} \\
	\end{array} } \right]
\label{eq:alpha7}
\end{equation}

\noindent A comparison of the \textcolor{model7}{top} and \textcolor{model6}{middle} 3 $\times$ 3 block matrices in Eq.~\ref{eq:alpha7} to the $\hat{L}_{z}$ operator (normalized by $\hbar$) in the $|l=1,m_{l}\rangle$ basis given by 
 
 \begin{equation}
 \hat{L}_{z} = \left[ {\begin{array}{ccc}
 	-1 & 0  & 0  \\
 	0 & 0  & 0  \\
 	0 & 0  & 1 \\
 	\end{array} } \right]
 \label{eq:alpha8}
 \end{equation}
 
 \noindent confirms that both block matrices have equivalent arrangements of matrix elements to $\hat{L}_{z}$ operator in the $|l=1,m_{l}\rangle$ basis, with projection factors $\alpha=-\frac{3}{2}$ and $\frac{1}{2}$ for the ground and first excited manifolds, respectively, in agreement with previous derivations utilizing group theory~\cite{kanamori,abragam,seitz1959,cowley}. \\
\indent As a final confirmation of the validity of the projection described by Eq.~\ref{eq:transform}, both $\hat{L}_{+}$ and $\hat{L}_{-}$ were projected onto the $|\phi\rm{_{CEF}}\rangle$ basis. Both $\hat{L}_{x}$ and $\hat{L}_{y}$ were then calculated using the following identities:
 
 \begin{equation}
 \hat{L}_{x} = \frac{\hat{L}_{+}+\hat{L}_{-}}{2}
 \label{eq:plus}
 \end{equation}   
 
 \noindent and
 \begin{equation}
 \hat{L}_{y} = \frac{\hat{L}_{+}-\hat{L}_{-}}{2i}.
 \label{eq:minus}
 \end{equation} 
 
 yielding:  
 
  \begin{equation}
 \hat{L}_{x} = \left[ {\begin{array}{ccc|ccc|c}
 	\textcolor{model7}{0} & \textcolor{model7}{1.1}  & \textcolor{model7}{0} & 0 & -1.4& 0 & 0 \\
\textcolor{model7}{1.1} & \textcolor{model7}{0}  & \textcolor{model7}{-1.1} & -1.4 & 0& -1.4 & 0 \\
\textcolor{model7}{0} & \textcolor{model7}{-1.1}  & \textcolor{model7}{0} & 0 & 1.4& 0 & 0 \\
\hline
0 & -1.4  & 0 & 	\textcolor{model6}{0} & 	\textcolor{model6}{0.4}& 	\textcolor{model6}{0} & -1.4 \\
-1.4 & 0  & 1.4 & 	\textcolor{model6}{0.4} & 	\textcolor{model6}{0} & 	\textcolor{model6}{0.4} & 0 \\
0 & -1.4  & 0 & 	\textcolor{model6}{0} & \textcolor{model6}{0.4}& \textcolor{model6}{0} & 1.4 \\
\hline
0  & 0  & 0 & -1.4 & 0& 1.4 & \textcolor{model2}{0}  \\
 	\end{array} } \right]
 \label{eq:alpha12}
 \end{equation}
  
    \begin{equation}
  \hat{L}_{y} = i\left[ {\begin{array}{ccc|ccc|c}
 	\textcolor{model7}{0} & \textcolor{model7}{1.1}  & \textcolor{model7}{0} & 0 & 0& 0 & 0 \\
\textcolor{model7}{-1.1} & \textcolor{model7}{0}  & \textcolor{model7}{-1.1} & -1.4 & 0& 1.4 & 0 \\
\textcolor{model7}{0} & \textcolor{model7}{1.1}  & \textcolor{model7}{0} & 0 & 1.4& 0 & 0 \\
\hline
0 & 1.4  & 0 & 	\textcolor{model6}{0} & 	\textcolor{model6}{0.4}& 	\textcolor{model6}{0} & -1.4 \\
-1.4 & 0  & -1.4 & 	\textcolor{model6}{-0.4} & 	\textcolor{model6}{0} & 	\textcolor{model6}{0.4} & 0 \\
0 & -1.4  & 0 & 	\textcolor{model6}{0} & \textcolor{model6}{-0.4}& \textcolor{model6}{0} & -1.4 \\
\hline
\xi  & 0  & 0 & 1.4 & 0& 1.4 & \textcolor{model2}{0}  \\
  	\end{array} } \right].
  \label{eq:alpha13}
  \end{equation}
  
\noindent Finally, by extracting the \textcolor{model7}{top} $3 \times 3$ block matrices, denoted by a prime, from the definitions of $\hat{L}_{z}$ (Eq.~\ref{eq:alpha7}), $\hat{L}_{x}$ (Eq.~\ref{eq:alpha12}) and $\hat{L}_{y}$ (Eq.~\ref{eq:alpha13}) and evaluating the commutator $[\hat{L}'_{x},\hat{L}'_{y}]$, one obtains 

\begin{equation}
[\hat{L}'_{x},\hat{L}'_{y}] = i\left[ {\begin{array}{ccc}
	\textcolor{model7}{1.5} & \textcolor{model7}{0}  & \textcolor{model7}{0}  \\
	\textcolor{model7}{0} & \textcolor{model7}{0}  & \textcolor{model7}{0}  \\
	\textcolor{model7}{0} & \textcolor{model7}{0}  & \textcolor{model7}{-1.5} \\
	\end{array} } \right] = i\hat{L}'_{z}.
\label{eq:alpha14}
\end{equation}
 
\indent By performing the commutator of all possible permutations of the projected components of the orbital angular momentum operator, it can be shown that the canonical commutation relations of angular momentum~\cite{abragam}, normalized by $\hbar$, 

\begin{equation}
[\hat{L}'_{x},\hat{L}'_{y}] = i\epsilon_{xyz}\hat{L}'_{z} 
\label{eq:alpha16}
\end{equation}    

\noindent are satisfied for the new $|\phi\rm{_{CEF}}\rangle$ basis.
 
 \label{sec:appendix_alpha} 
 
\subsection{Calculation of Projected Land\'{e} $g$-Factor $g'_{J}$}\label{sec:gj}

\hspace*{5.00mm} Recall from first-order perturbation theory~\cite{buyers}, the field splitting of the Co$^{2+}$ spin-orbit multiplets is described by the perturbative Hamiltonian $\hat{\mathcal{H}}_{m}$ given by 

\begin{equation}
\hat{\mathcal{H}}_{m} = \mu_{B}(g_{L}\hat{\mathbf{L}} + g_{S}\hat{\mathbf{S}})\cdot\mathbf{H},
\label{eq:gfactor_1}
\end{equation} 

\noindent where $g_{L}$ and $g_{S}$ denote orbital and spin $g$-factors, respectively. For the particular case of the $d$-block metal Co$^{2+}$, both orbital and spin $g$-factors are taken to be the electron's $g$-factors, equal to approximately 1 and 2, respectively, simplifying Eq.~\ref{eq:gfactor_1} to 

\begin{equation}
\hat{\mathcal{H}}_{m} = \mu_{B}(\hat{\mathbf{L}} + 2\hat{\mathbf{S}})\cdot\mathbf{H}. 
\label{eq:gfactor_2}
\end{equation}

\noindent Since an effective total angular momentum $\hat{\mathbf{j}}\rm{_{eff}}$ was defined with the projected orbital angular momentum operator $\hat{\mathbf{l}}$ with $l=$1, then the perturbative Hamiltonian in Eq.~\ref{eq:gfactor_2} becomes

\begin{align}
\hat{\mathcal{H}}_{m} &=  \mu_{B}(\alpha\hat{\mathbf{l}} + 2\hat{\mathbf{S}})\cdot\mathbf{H} \nonumber \\
&= {} g'_{J}\mu_{B}\hat{\mathbf{j}}\cdot\mathbf{H}, 
\label{eq:gfactor_3}
\end{align} 

\noindent for a particular effective spin-orbit $j\rm{_{eff}}$ manifold. Eq.~\ref{eq:gfactor_3} incorporates an orbital angular momentum operator $\hat{\mathbf{L}}$ that has been projected onto $\hat{\mathbf{l}}$ \emph{via} a projection factor $\alpha$, and a projected Land\'{e} $g$-factor $g'_{J}$. A comparison between Eqs.~\ref{eq:gfactor_2} and~\ref{eq:gfactor_3} suggests that the Land\'{e} $g$-factor --- a fundamental proportionality constant that can be derived directly from the Wigner-Eckart theorem~\cite{sarte18:98} --- defined as  

\begin{align}
g_{J} &= 1\left\{\frac{J(J+1)-S(S+1)+L(L+1)}{2J(J+1)}\right\} \nonumber \\
&\quad {} + 2\left\{\frac{J(J+1)+S(S+1)-L(L+1)}{2J(J+1)}\right\}
\label{eq:g_j} 
\end{align} 

\noindent for the original non-projected perturbative Hamiltonian in Eq.~\ref{eq:gfactor_2} assumes the form

\begin{equation}
g'_{J} = \frac{(2+\alpha)j(j+1) - (2-\alpha)l(l+1) + (2-\alpha)S(S+1) }{2j(j+1)}.
\label{eq:g_j_prime}
\end{equation}

\noindent As required, Eq.~\ref{eq:g_j_prime} reduces to Eq.~\ref{eq:g_j} if $\alpha=1$. By inserting the values of $S=\frac{3}{2}$ to reflect the high spin $d^{7}$ electron configuration in ideal octahedral coordination, $l=1$ to reflect the ground state crystal field manifold and the associated projection factor $\alpha=-\frac{3}{2}$, the projected Land\'{e} $g$-factor of $\frac{13}{3}$ is obtained for the $j\rm{_{eff}}=\frac{1}{2}$ ground state spin-orbit manifold~\cite{buyers}.  

\subsection{Calculation of Spin Angular Momentum Operator Projection Factor $\alpha'$}
 
\indent As discussed in previous work~\cite{ross,sarte18:98,buyers,cowley,kanamori,cowley_Co} on other systems whose magnetism is based on Co$^{2+}$ in octahedral coordination, multiple projections of different angular momentum operators are necessary to consolidate the measured low temperature magnetic excitations and the theoretical framework for a $j\rm{_{eff}}=\frac{1}{2}$ ground state. One method~\cite{fiete} for such  projections was presented in Appendix~\ref{sec:projection} and involved the use of linear transformations in the matrix representation of operators. Although powerful, this method relies on access to computation software and quickly becomes tedious as the dimension of the Hilbert space of interest increases. For the purposes of completion, we present an alternative method to project angular momentum operators onto a particular $j\rm{_{eff}}$ manifold. This method consists of a special case of the Wigner-Eckart theorem~\cite{projection,abragam}, called the projection theorem, given by

\begin{equation}
\hat{\mathbf{\mathcal{O}}} = \alpha'\hat{\mathbf{j}}{\rm{_{eff}}} = \frac{\langle \hat{\mathbf{\mathcal{O}}} \cdot \hat{\mathbf{j}}\rm{_{eff}} \rangle}{j(j+1)}\hat{\mathbf{j}}\rm{_{eff}}, 
\label{eq:ap5}
\end{equation} 

\noindent describing the projection of an angular momentum operator $\hat{\mathbf{\mathcal{O}}}$ onto an effective total angular momentum operator $\hat{\mathbf{j}}\rm{_{eff}}$ \emph{via} a projection factor $\alpha'$. As introduced in Appendix~\ref{sec:appendix_alpha}, the operator $\hat{\mathbf{j}}\rm{_{eff}} = \hat{\mathbf{l}} + \hat{\mathbf{S}}$ denotes an effective total angular momentum operator that utilizes a projection of an orbital angular momentum operator $\hat{\mathbf{L}}$ with $L=3$ onto a fictitious orbital angular momentum operator $\hat{\mathbf{l}}$ with $l=1$ \emph{via} $\alpha$. \\
\indent For illustrative purposes, let $\hat{\mathbf{\mathcal{O}}}$ be the spin angular momentum operator $\hat{\mathbf{S}}$. The numerator of the projection factor $\alpha'$ in Eq.~\ref{eq:ap5} can be simplified by first using the distributive property of the inner product 

\begin{equation}
\hat{\mathbf{S}} \cdot \hat{\mathbf{j}}{\rm{_{eff}}} = \hat{\mathbf{S}}\cdot (\hat{\mathbf{l}} + \hat{\mathbf{S}}) = \hat{S}^{2} + \hat{\mathbf{l}} \cdot \hat{\mathbf{S}}.
\label{eq:ap6}
\end{equation}

\noindent The inner product $\hat{\mathbf{l}}\cdot\hat{\mathbf{S}}$ on the RHS of Eq.~\ref{eq:ap6} can be simplified to

\begin{equation}
\hat{\mathbf{l}} \cdot \hat{\mathbf{S}} = \frac{1}{2}\left[(\hat{j}{\rm{_{eff}}})^{2}-\hat{l}^{2}-\hat{S}^{2} \right],
\label{eq:ap7}
\end{equation}

\noindent since the inner product of $\hat{\mathbf{j}}\rm{_{eff}}$ with itself is equal to 

\begin{equation}
(\hat{j}{\rm{_{eff}}})^{2}  = (\hat{\mathbf{l}} + \hat{\mathbf{S}}) \cdot (\hat{\mathbf{l}} + \hat{\mathbf{S}}) = \hat{l}^{2} + \hat{S}^{2} + 2 \hat{\mathbf{l}} \cdot \hat{\mathbf{S}}.
\label{eq:ap8}
\end{equation}

\noindent Combining Eqs.~\ref{eq:ap6} and~\ref{eq:ap7}, the numerator of $\alpha'$ in Eq.~\ref{eq:ap5} becomes  

\begin{equation}
\langle\hat{\mathbf{S}} \cdot \hat{\mathbf{j}}{\rm{_{eff}}}\rangle = S(S+1) + \frac{1}{2}\left[j(j+1) -l(l+1) - S(S+1) \right],
\label{eq:ap9}
\end{equation}

\noindent where $j\rm{_{eff}}$ was relabeled as $j$. Inserting Eq.~\ref{eq:ap9} into Eq.~\ref{eq:ap5}, one obtains
\begin{equation}
\hat{\mathbf{S}} = \frac{S(S+1) + \frac{1}{2}\left[ j(j+1) -l(l+1) - S(S+1) \right]}{j(j+1)}\hat{\mathbf{j}}\rm{_{eff}},
\end{equation}

\noindent which can be simplified algebraically to 

\begin{equation}
\hat{\mathbf{S}} = \left\{\frac{1}{2} + \frac{S(S+1) -l(l+1)}{2j(j+1)} \right\}\hat{\mathbf{j}}\rm{_{eff}}. 
\label{eq:ap11}
\end{equation}

\noindent Finally, by inserting the aforementioned values of $S=\frac{3}{2}$, $l=1$ and $j\equiv j\rm{_{eff}}=\frac{1}{2}$ for high spin Co$^{2+}$, Eq.~\ref{eq:ap11} simplifies to 

\begin{equation}
\hat{\mathbf{S}}=\frac{5}{3}\hat{\mathbf{j}}\rm{_{eff}}. 
\label{eq:ap10}
\end{equation}

\noindent A comparison between Eqs.~\ref{eq:ap5} and \ref{eq:ap10} indicates that the projection factor $\alpha'$ of the spin orbital angular momentum operator is $\frac{5}{3}$ for the $j=\frac{1}{2}$ ground state spin-orbit manifold~\cite{sarte18:98}. It can be shown~\cite{kanamori} that one obtains the same value of $\alpha'$ employing the method outlined in Appendix~\ref{sec:projection} with the transformation matrix $\mathcal{C}$ defined as the eigenvectors of the spin-orbit Hamiltonian $\hat{\mathcal{H}}_{SO} = \alpha\lambda\hat{\mathbf{l}}\cdot\hat{\mathbf{S}}$, where $\alpha=-\frac{3}{2}$ as derived in Appendix~\ref{sec:projection}, $\lambda=-16$~meV as measured by Cowley \emph{et al.}~\cite{cowley}, and $\hat{\mathbf{l}}$ is a fictitious orbital angular momentum operator with an eigenvalue $l$=1 as discussed above and in the main text.  

\section{Derivation of the Powder-Averaged First Moment Sum Rule of Neutron Scattering}

The first moment sum rule of neutron scattering~\cite{hohenberg} states that 

\begin{equation}
\langle E \rangle (\mathbf{Q}) = -\frac{2}{3}\sum\limits_{i,j}n_{ij}J_{ij}\langle \hat{\mathbf{S}}_{i} \cdot \hat{\mathbf{S}}_{j}\rangle(1-\cos(\mathbf{Q}\cdot \mathbf{d_{ij}})), 
\label{eq:1}
\end{equation}

\noindent where $J_{ij}$, $n_{ij}\langle \hat{\mathbf{S}}_{i} \cdot \hat{\mathbf{S}}_{j}\rangle$, $\mathbf{d}_{ij}$ denote the exchange constant, spin-spin correlator and displacement vector between spins $i$ and $j$, respectively. Applying the definition of the powder average~\cite{dimer}, Eq.~\ref{eq:1} becomes

\begin{equation}
S(|\mathbf{Q}|,E) = \int d\Omega_{\hat{\mathbf{Q}}} \frac{S(\mathbf{Q},E)}{4\pi},
\label{eq:powderaverage}
\end{equation}

\noindent and utilizing the property of linearity of the integral, one obtains

\begin{equation}
-\frac{\mathcal{B}_{ij}}{12\pi}\int_{0}^{\pi}\int_{0}^{2\pi}(1-\cos(|\mathbf{Q}||\mathbf{d}_{ij}|\cos\theta))d\phi\sin\theta d\theta,
\label{eq:2}
\end{equation}

\noindent where $\mathcal{B}_{ij}$ denotes $2n_{ij}J_{ij}\langle \hat{\mathbf{S}}_{i} \cdot \hat{\mathbf{S}}_{j}\rangle$ for a particular $ij$ pair type. Using the substitution of $x~=~|\mathbf{Q}||\mathbf{d}_{ij}|\cos\theta$ in Eq.~\ref{eq:2}, one obtains

\begin{equation} \frac{\mathcal{B}_{ij}}{3}\left(\frac{1}{4\pi}\right)\int_{|\mathbf{Q}||\mathbf{d}_{ij}|}^{-|\mathbf{Q}||\mathbf{d}_{ij}|}\int_{0}^{2\pi}(1-\cos(x))d\phi\frac{dx}{|\mathbf{Q}||\mathbf{d}_{ij}|}.
\label{eq:3}
\end{equation} \newline 

\noindent Employing the linearity property of the integral, the first term in Eq.~\ref{eq:3} is reduced to

\begin{equation}
\frac{\mathcal{B}_{ij}}{3}\left(\frac{1}{4\pi}\right)\int_{|\mathbf{Q}||\mathbf{d}_{ij}|}^{-|\mathbf{Q}||\mathbf{d}_{ij}|}\int_{0}^{2\pi}\frac{d\phi dx}{|\mathbf{Q}||\mathbf{d}_{ij}|} = -\frac{\mathcal{B}_{ij}}{3}.
\label{eq:first}
\end{equation}

\noindent The second term in Eq.~\ref{eq:3} becomes 

\begin{equation}
-\frac{\mathcal{B}_{ij}}{3}\left(\frac{1}{4\pi}\right)\int_{|\mathbf{Q}||\mathbf{d}_{ij}|}^{-|\mathbf{Q}||\mathbf{d}_{ij}|}\int_{0}^{2\pi}\cos(x)\frac{d\phi dx}{|\mathbf{Q}||\mathbf{d}_{ij}|},
\label{eq:second}
\end{equation}

\noindent which can be simplified by first integrating out $d\phi$,

\begin{equation}
-\frac{\mathcal{B}_{ij}}{3}\left(\frac{2\pi}{4\pi}\right)\int_{-|\mathbf{Q}||\mathbf{d}_{ij}|}^{|\mathbf{Q}||\mathbf{d}_{ij}|}\cos(x)\frac{dx}{|\mathbf{Q}||\mathbf{d}_{ij}|},
\label{eq:second_2}
\end{equation}

\noindent which is equal to 

\begin{equation}
-\frac{\mathcal{B}_{ij}}{3}\left(\frac{2\pi}{4\pi}\right)(\sin(-|\mathbf{Q}||\mathbf{d}_{ij}|)-\sin(|\mathbf{Q}||\mathbf{d}_{ij}|)).
\label{eq:second_3}
\end{equation}

\noindent Since sine is an odd function, Eq.~\ref{eq:second_3} reduces to

\begin{equation}
\frac{\mathcal{B}_{ij}}{3}\left(\frac{\sin(|\mathbf{Q}||\mathbf{d}_{ij}|}{|\mathbf{Q}||\mathbf{d}_{ij}|}\right).
\label{eq:second_4}
\end{equation}

\noindent Combining both terms, one obtains the final expression for the $|\mathbf{Q}|$-dependence of the powder-averaged first moment as 

\begin{equation} 
\langle E \rangle(|\mathbf{Q}|) =  -\frac{\mathcal{B}_{ij}}{3}\left(1-\frac{\sin(|\mathbf{Q}||\mathbf{d}_{ij}|}{|\mathbf{Q}||\mathbf{d}_{ij}|} \right).
\label{eq:final}
\end{equation}

\noindent The expression in Eq.~\ref{eq:final} pertains to one particular $ij$ pair-type. Utilizing the linearity property of the integral and replacing $\mathcal{B}_{ij}$ by its definition, one can recover the sum from Eq.~\ref{eq:1}, 

\begin{equation} \langle E \rangle (\mathbf{Q}) =-\frac{2}{3}\sum\limits_{i,j}n_{ij}J_{ij}\langle \hat{\mathbf{S}}_{i} \cdot \hat{\mathbf{S}}_{j}\rangle\left(1-\frac{\sin(|\mathbf{Q}||\mathbf{d}_{ij}|}{|\mathbf{Q}||\mathbf{d}_{ij}|} \right),
\label{eq:final2}
\end{equation}

\noindent corresponding to Eq.~\ref{eq:firstmoment2} in the main text.

\renewcommand{\figurename}{Figure S}
\renewcommand{\topfraction}{0.85}
\renewcommand{\bottomfraction}{0.85}
\renewcommand{\textfraction}{0.1}
\renewcommand{\floatpagefraction}{0.75}


\bibliography{references_CoV3O8}

\end{document}